\documentclass[aps,prf,notitlepage,amsmath,amssymb,floatfix,longbibliography,12pt,tightenlines,superscriptaddress]{revtex4-1}

\usepackage{graphicx}
\usepackage{dcolumn}
\usepackage{bm}
\usepackage[percent]{overpic}
\usepackage[usenames,dvipsnames,svgnames,table]{xcolor}
\usepackage{subfigure}
\usepackage{ulem}
\usepackage{filecontents}
\usepackage{float}
\usepackage{hyperref}
\usepackage{mathtools}
\usepackage{verbatim}
\usepackage{natbib}
\hypersetup{colorlinks=true, citecolor=blue, linkcolor=blue}

\newcommand\BV{{Brunt-V\"ais\"al\"a frequency}} 

\begin{document}
	
	\title{Turbulence generation by large-scale extreme vertical drafts and the modulation of local energy dissipation in stably stratified geophysical flows}
	
	\begin{abstract}
		We observe the emergence of strong vertical drafts in direct numerical simulations of the Boussinesq equations in a range of parameters of geophysical interest. These structures, which appear intermittently in space and time, generate turbulence and enhance kinetic and potential energy dissipation, providing a possible explanation for the observed variability of the local energy dissipation in the bulk of oceanic flows, and the modulation of its probability distribution function. We show how, due to the extreme drafts, in runs with Froude numbers observable in geophysical scenarios, roughly $10\%$ of the domain flow can account for up to $50\%$ of the global volume dissipation, reminiscent of estimates based on oceanic models.
	\end{abstract}
	
	\author{Raffaele Marino}
	\affiliation{Laboratoire de M\'ecanique des Fluides et d'Acoustique, CNRS, \'Ecole Centrale de Lyon, Universit\'e Claude Bernard Lyon 1, INSA de Lyon, F-69134 \'Ecully, France.}
	\author{Fabio Feraco}
	\affiliation{Laboratoire de M\'ecanique des Fluides et d'Acoustique, CNRS, \'Ecole Centrale de Lyon, Universit\'e Claude Bernard Lyon 1, INSA de Lyon, F-69134 \'Ecully, France.}
	\affiliation{Dipartimento di Fisica, Universit\`a della Calabria, Italy.}
	\author{Leonardo Primavera}
	\affiliation{Dipartimento di Fisica, Universit\`a della Calabria, Italy.}
	\author{Alain Pumir}
	\affiliation{Laboratoire de Physique, UMR5672, \'Ecole Normale Sup\'erieure de Lyon and CNRS, 46 All\'ee d'Italie, F-69007, Lyon, France.}
	\author{Annick Pouquet}
	\affiliation{Laboratory for Atmospheric and Space Physics, University of Colorado, Boulder, CO 80309, USA.}
	\author{Duane Rosenberg}
	\affiliation{288 Harper St. Louisville, CO 80027 USA.}
	\author{Pablo D. Mininni}
	\affiliation{Departamento de F\'{\i}sica, Facultad de Ciencias Exactas y Naturales, Universidad de Buenos Aires, and IFIBA, CONICET, Buenos Aires 1428, Argentina.}
	
	\maketitle
	
	\section{Introduction}
	The combination of turbulent eddies and waves, due to stratification and rotation, leads to the formation of surprising features in geophysical flows, from oval structures and jets on Jupiter \cite{marcus_93}, to hurricanes and tornadoes in our atmosphere \cite{emanuel_05}, or strong currents and dual energy cascades in the ocean \cite{scott_05,klymak_08,klein_08,marino_13i,pouquet_13,marino_14,marino_15p,pouquet_17,pouquet_19,pouquet_19e,xie1,xie2}. The interplay between such structures is not fully understood, in particular their dependence on control parameters such as the Reynolds and Froude numbers \cite{marino_15w,pouquet_18j,buhler,herbert_16} (see definitions below). But it is known that extreme events associated with these structures can play an important role in the dynamics and dissipation. For example, sudden and significant enhancements of the vertical velocity (hereafter ``drafts") have been observed in geophysical flows, in the planetary boundary layer \cite{lenschow_12,mahrt1,mahrt2}, as well as far from boundaries in the mesosphere and lower thermosphere (MLT) \cite{liu_07,chau_21} and in the ocean at depths close to that of the mixed layer \cite{dasaro}. Moreover, in the ocean, vast regions of rather constant energy dissipation $\varepsilon_0$ are observed together with local patches of turbulence dissipating $10^2 \varepsilon_0$ to $10^4 \varepsilon_0$, as in the vicinity of the Hawaiian ridge \cite{klymak_08} or of the Puerto-Rico trench \cite{vanharen_16j}, with large values of the vertical velocity \cite{dasaro,capet}. Similarly, in frontal structures, gradients of tracers such as pollutants are characterized by very large fluctuations, with a kurtosis (the normalized fourth moment of the distribution) reaching values of several hundreds~\cite{klymak_15}. All these extreme events are characterized by non-Gaussian statistics.
	
	The occurrence of many of these extreme events is associated with turbulence. Unexpectedly, early seminal studies 
	of stratified flows \cite{Riley_1981, Herring_1989, Winters_1995, Metais_1996} revealed that stably stratified turbulence, as found, e.g., at intermediate scales in the nocturnal atmosphere and in the oceans, is more complex than quasi-geostrophic dynamics and very different from homogeneous and isotropic turbulence (HIT). This is evidenced, e.g., by the formation of anisotropic horizontal structures and their effect on turbulent transport \cite{Kimura_1996, Billant_2000}, or by the complex {structure of the spectra, characterizing the kinetic and potential energy of the flow} \cite{Riley_2003, Lindborg_2006, marino_14}. Stably stratified turbulence is also dependent on the Reynolds number \cite{Laval_2003}, {the possible flow regimes being
		controlled by the} product of the Reynolds and the squared Froude number \cite{ Bartello_2013}, the so-called buoyancy Reynolds number (see \cite{ivey_08} for a review, and for an alternate definition to the one used in this work). 
	These results led to the emergence of a physical picture for strongly stratified turbulence, in which an anisotropic and forward energy cascade is associated with highly anisotropic vortical structures, and with the development of breakdown events on small scales where the flow becomes super-critical, feeding into local patches of more isotropic dynamics (see, e.g., \cite{Riley_2003, Brethouwer_2007, davidson, pouquet_19p}). Concerning extreme events, it is known that stably stratified turbulence displays
	intermittency
	\cite{rorai2014,deBruyn_2015,feraco2018,Smyth_2019,feraco2021}, but only recently {it was} found that the amount of {large-scale}  intermittency (to distinguish it from the classical small-scale intermittency considered in several studies \cite{deBruyn_2015}) depends {sharply} on the Froude number, with some regimes being in fact more intermittent than HIT, and displaying extreme values of the large-scale vertical flow velocity \cite{rorai2014, feraco2018, feraco2021},
	as also observed for example in reanalysis of climate data \cite{petoukhov_08, sardeshmukh_15}.
	These latter events are of a different nature than the small-scale extreme events, as they involve directly the velocity instead of velocity gradients. However, their effect on the energy dissipation, and how these extreme vertical drafts interact with and affect the turbulence, remain unclear.
	
	Recent numerical work based on the Boussinesq equations has confirmed that the vertical component of the velocity field ($w$) indeed exhibits a large-scale intermittent behavior, in both space and time, for values of Froude numbers of geophysical interest \cite{feraco2018}, with a connection existing between large- and small-scale intermittency in stratified turbulent flows \cite{feraco2021}. In particular, direct numerical simulations (DNSs) of stably stratified turbulent flows were found to systematically develop powerful vertical drafts that make the statistics of $w$ strongly non-Gaussian in the energy-containing eddies. This is interpreted as a result of the interplay between gravity waves and turbulent motions, and it occurs in a resonant regime of the governing parameters where vertical velocities are enhanced much faster than in the analogous HIT case \cite{rorai2014,feraco2018,feraco2021}. It can also be understood as a result of complex phase-space dynamics in a reduced model for the velocity and temperature gradients \cite{sujovolsky_20}.
	The present study establishes that extreme drafts emerge in a recurrent manner in stratified flows, producing localized turbulence, and ultimately bursts of dissipation at small-scale,
	{as observed for example in oceanic data and DNSs \cite{salehipour_16} (see also \cite{Smyth_2019}).}
	
	\section{Simulations and parameters}
	We performed a series of DNSs in a triply-periodic domain of side $2\pi L_0$ with $512^3$ grid points for up to $t \approx 500 \tau_{NL}$ ($\tau_{NL}=L/U$ being the turnover time, $U$ and $L$ the flow characteristic velocity and integral scale respectively in units of a simulation unit length $L_0$ and a unit velocity $U_0$ with {$T_0=L_0/U_0$}). {In these units, in all the simulations considered below, the velocity $u$ is
		$U\approx 1$ to $1.5$, and the flow integral scale is $L\approx 2\pi/\overline{k}_f \approx 2.5$ in all cases, with $\overline{k}_f$ the mean forced wavenumber; for practical purposes the typical velocity $U$ and length $L$ can be considered $\mathcal{O}(1)$.} Some of these flows were analyzed in \cite{feraco2018} over a limited time span (up to $\approx 25 \tau_{NL}$), while other simulations analyzed herein are new. It was shown in \cite{feraco2018,feraco2021} 
	that the kurtosis of $w$ could reach high values in a narrow regime of parameters around a Froude number $\textrm{Fr}=U/(LN)\approx 0.08$, with $N=[-(g/\rho_0)\partial_z\overline\rho]^{1/2}$ the \BV, $g$ the gravitational acceleration, $\rho_0$ the mean density, and $\overline\rho$ the background linear density profile. Simulation parameters are listed in Table \ref{t:runs}. Runs P3 to P5, with $5 \leq NT_0 \leq 8$, correspond to the resonant regime mentioned above. In all the simulations with $N\neq 0$,  $NT_0$ varies from $1.5$ to $23.5$. As a reference, a typical velocity in the ocean of $U_0=0.1$ ms$^{-1}$, and a unit length of $1$ km (thus for a computational domain of $\approx 6.3$ km), results in $N \approx 2\times 10^{-3}$ s$^{-1}$ (for run P7) to $N \approx 10^{-4}$ s$^{-1}$ (for run P2). These values are reasonable for oceanic scales and situations in which the hydrostatic approximation breaks down \cite{Vallis}.
	
	The Boussinesq equations for the velocity $\textbf u$ and the density fluctuation $\rho'$ around the stable linear background are
	\begin{eqnarray}\label{bequationV}
	\partial_t \textbf u+(\textbf u \cdot\nabla)\textbf u 
	&=& - \nabla (p/\rho_0) - (g/\rho_0) \rho' \, \hat{z} + \textbf F_u +\nu\nabla^2\textbf u, \\
	\partial_t\rho' +\textbf u\cdot\nabla\rho' &=& (\rho_0 N^2/g) w + \kappa\nabla^2\rho' \ ,
	\label{bequationT} \end{eqnarray}
	where $\kappa$ is the thermal diffusivity and $\nu$ the kinematic viscosity, with $\kappa=\nu$ for all runs. Both values, in units of $L_0U_0$ in Table \ref{t:runs}, are not realistic for geophysical flows and come as a result of computational constraints, and should thus be considered as effective transport coefficients. In spite of this, in the simulations presented here, turbulence is strong enough for the typical nonlinear dynamics observed in the atmosphere and in the oceans to develop. We take the convention that the vertical coordinate, $z$, points upwards, and gravity downwards. The total fluid density is $\rho = \overline\rho(z) + \rho'$, with $\langle \rho' \rangle = 0$, and uniform $\partial_z \overline\rho <0$, expressing that the background density decreases linearly with $z$ ($\langle \overline\rho(z) \rangle = \rho_0$). These equations can also be written using a scaled density fluctuation $\zeta = g\rho'/(N\rho_0)$, with units of velocity $U_0$, as
	\begin{eqnarray}\label{bequationV}
	\partial_t \textbf u+(\textbf u \cdot\nabla)\textbf u 
	&=& - \nabla p' - N\zeta\hat{z} + \textbf F_u +\nu\nabla^2\textbf u, \\
	\partial_t\zeta+\textbf u\cdot\nabla\zeta &=& Nw + \kappa\nabla^2\zeta \ ,
	\label{bequationT} \end{eqnarray}
	where $p'=p/\rho_0$. This latter form is convenient as the kinetic and potential energy (per unit mean density $\rho_0$) are {then}
	given respectively by $E_V = \int u^2/2 \, dV$ and $E_P = \int \zeta^2/2 \, dV$.
	
	\begin{table}
		\begin{ruledtabular}
			\begin{tabular}{c c c c c c c c c c}
				\hfill
				Run &P1& P2& P3& P4& P5& P6& P7& P8& P9\\
				\hline
				\hline
				$\textrm{Re}$ $[\times 10^3]$ & 2.4 & 2.6 & 3.6 & 3.8 & 3.8 & 3.8 & 3.8 & 1.2 & 0.8 \\
				$\textrm{Fr}$ $[\times 10^{-1}]$ & $\infty$ & 2.8 & 1.1 & 0.81 & 0.76 & 0.3 & 0.26 & 0.76 & 0.71 \\
				$\textrm{R}_B$ & $\infty$ & 206 & 43.8 & 24.8 & 22.1 & 3.4 & 2.6 & 6.8 & 4.2 \\
				\hline
				$\nu [\times 10^{-3} L_0 U_0]$ & 1.5 & 1 & 1 & 1 & 1 & 1 & 1 & 3 & 4 \\
				$N$ $[U_0/L_0]$ & 0 & 1.5 & 5 & 7.37 & 8 & 20 & 23.5 & 7.37 & 7.37 \\
				$U$ $[U_0]$ & 1.4 & 1.0 & 1.4 & 1.5 & 1.5 & 1.5 & 1.5 & 1.4 & 1.3 \\
				$t_{tot}/\tau_{NL}$ & 30  & 55 & 103 & 452 & 406 & 91 & 62 & 526 & 422 \\ 
			\end{tabular}
		\end{ruledtabular}
		\caption{Parameters of the runs, with $\textrm{Re}$, $\textrm{Fr}$, and $\textrm{R}_B$ respectively the Reynolds, Froude, and buoyancy Reynolds numbers, and $\nu$, $N$, $U$, and $t_{tot}/\tau_{NL}$ the kinematic viscosity, the Brunt-V\"ais\"al\"a frequency, {the flow typical (r.m.s.) velocity,} and the temporal extension of the runs in units of $\tau_{NL}$. In all runs $L\approx 2.5 L_0$, and $\tau_{NL} = L/U$.}
		\label{t:runs}
	\end{table}
	
	The Reynolds and buoyancy Reynolds numbers are $\textrm{Re}=UL/\nu$ and $\textrm{R}_B = \textrm{Re} \, \textrm{Fr}^2$. $\textrm{R}_B$ is a measure of the relative strength of the buoyancy to the dissipation, 
	and allows for the identification of three regimes: one controlled by gravity waves ($\textrm{R}_B \le 10$), a transitional regime, and another dominated by turbulence ($\textrm{R}_B \ge 10^2$) \cite{Bartello_2013, feraco2018, pouquet_18j}. $\textrm{Fr}$, $\textrm{Re}$, and $\textrm{R}_B$ are computed for each run close to the peak of dissipation. A HIT run is also performed. The flows evolve under the action of a random forcing ${\textbf F}_u$, with constant amplitude, delta-correlated in time, isotropic in Fourier space, and centered on a spherical shell of wavenumbers $ 2 \le k_f L_0 \le3$.
	Simulations were performed with the GHOST code (Geophysical High Order Suite for Turbulence), a highly parallelized pseudo-spectral framework that hosts a variety of solvers \cite{mininni11, fontana_20, rosenberg_20} to study anisotropic classic and quantum fluids, as well as plasmas.
	
	\begin{figure}[t!]
		\begin{center}
			\includegraphics[trim={0 .8cm 0 .1cm},clip,width=13cm]{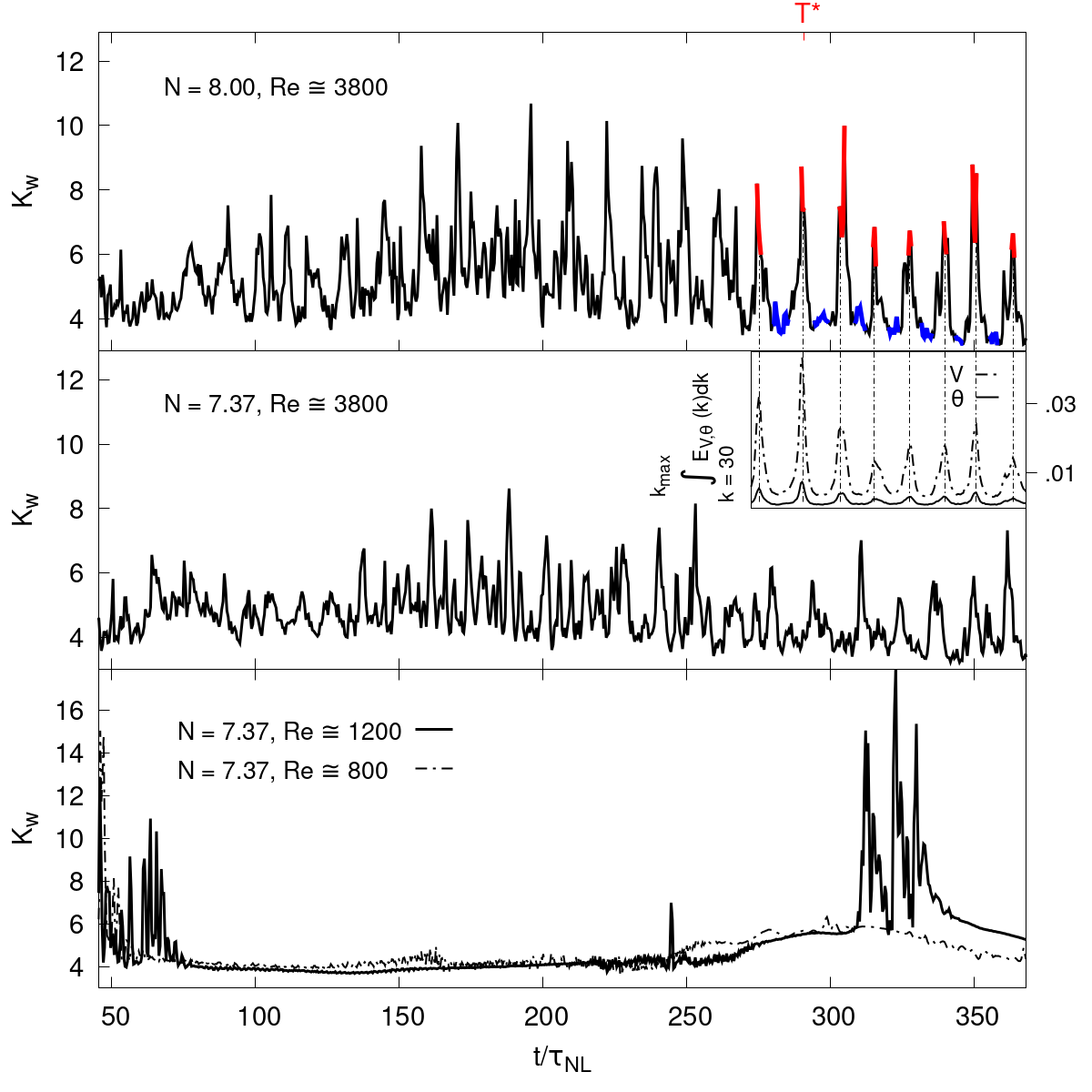}
		\end{center}
		\caption{Kurtosis $K_w$ of the vertical velocity $w$ as a function of time, in units of the turn-over time $\tau_{NL}$, for runs $P5$ (top panel), $P4$ (middle panel), as well as (bottom panel) $P8$ (solid line) and $P9$ (dashed line). Red and blue identify respectively the portions of the signal across relative maxima and minima used to compute the spectra in Fig.~\ref{Fig2}. $T^*\approx 290 \tau_{NL}$ is the reference time used in Figs. \ref{Fig3}. The inset in the middle panel shows the small-scale kinetic (V) and potential (P) energies at late times for run P5, obtained through the integration of the corresponding spectra from $k=30$ to $k_{max}\simeq512/3$.}
		\label{Fig1}
	\end{figure}
	
	\section{Generation of turbulence by extreme vertical drafts}
	The operative definition as well as the identification of the drafts in the DNSs performed in this study are obtained in terms of the statistical moments of the vertical components of the Eulerian velocity $w$. Specifically, we consider as drafts regions where the vertical velocity ($w$) is significantly larger than $3$ standard deviations ($\sigma_w$). The presence of the drafts in the simulations is then confirmed through the analysis of the temporal evolution of the kurtosis of the vertical velocity, $K_w=\langle (w - \langle{w}\rangle)^4 \rangle /\langle (w - \langle{w})\rangle^2 \rangle^2 $, computed as a function of time from the DNSs. We recall that the kurtosis of a Gaussian distribution is $K = 3$, so values of $K_w>3$ are indicative of the presence of extreme events in $w$. Fig.~\ref{Fig1} displays $K_w$ vs. $t/\tau_{NL}$ for runs P$4$, P$5$, P$8$, and P$9$. The Froude number for these runs is close to $\textrm{Fr} \approx 0.08$, for which $K_w$ was found to be maximum from calculations at $\textrm{Re} \approx 3800$ in~\cite{feraco2018}. 
	It is worth mentioning that the values of $\textrm{Fr}$ and $\textrm{R}_B$ of the main DNSs considered in this study are compatible with estimates of these parameters for some regions of the atmosphere and of quiet parts of the ocean interior.  
	To characterize how these extreme drafts affect the flow energetics,  we accumulate the statistics for several hundreds $\tau_{NL}$ (see Table \ref{t:runs}). Together with the investigation of this global quantity, our study will later consider statistical tools which are more local, either in Fourier or in physical space.
	
	Starting from the top panel of Fig.~\ref{Fig1} we observe that, for $N=8.0 U_0/L_0$, the flow is characterized by strong spikes of the kurtosis reaching values as high as $K_w \approx 11$, and separated by short time intervals with values of $K_w$ close to the Gaussian reference. The temporal analysis of the high-order spatial statistics allows us to conclude that in the  presence of large-scale intermittent drafts not only the flow is non-homogeneous due to the irregular emergence of these structures, but  furthermore, global properties of the flow exhibit wide fluctuations. A similar situation has been observed in other flows, such as turbulent homogeneous shear {flows~\cite{pumir96,sekimoto16}.} The fluctuations of $K_w$ are smaller for $N=7.37 U_0/L_0$ (Fig.~\ref{Fig1}, middle panel), which is still close to $\textrm{Fr} \approx 0.08$ and also appears to show fluctuations of large scale quantities. Keeping $N=7.37 U_0/L_0$ but lowering the Reynolds number to $\textrm{Re} \approx 1200$ changes drastically the dependence of $K_w$ on $t/\tau_{NL}$: it becomes a smooth signal interrupted by sporadic bursts. The signal is almost completely smooth and stationary for $\textrm{Re} \approx 800$ (Fig.~\ref{Fig1}, bottom panel). Since $\textrm{Fr}$ for the three runs in Fig.~\ref{Fig1} is roughly the same, this transition appears therefore to be
	led by the buoyancy Reynolds number $R_B$.
	
	\begin{figure}[b!]
		\includegraphics[width=13cm]{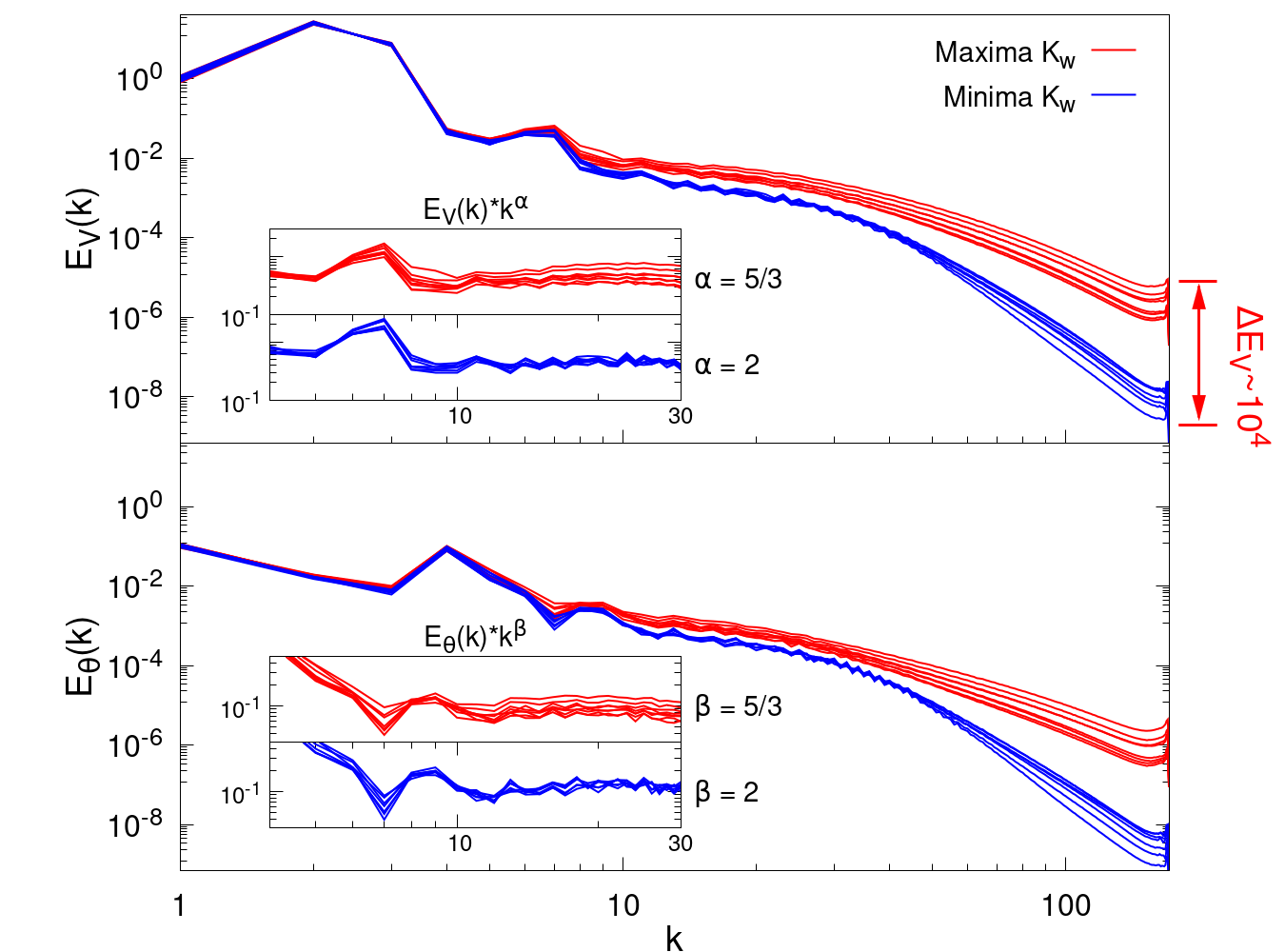}
		\caption{Kinetic (top) and potential (bottom) energy spectra computed for run $P5$ at the relative maxima (red) and minima (blue) of $K_w$ (see Fig.~\ref{Fig1}). Insets show $\alpha$ or $\beta$--compensated spectra respectively for $E_V$ or $E_P$, with $\alpha, \beta$ either $5/3$ or $2$.}
		\label{Fig2}
	\end{figure}
	
	To investigate how strong vertical drafts affect the generation of turbulence, we performed a systematic study of the isotropic kinetic and potential energy spectra, respectively $E_V(k)$ and $E_P(k)$. The results indicate a substantial enhancement of the power spectral density (PSD) in the small scales when $K_w$ is maximal. This can be clearly inferred from the spectra computed at local maxima and minima of the kurtosis $K_w$, as shown in Fig.~\ref{Fig2} for run $P5$ (corresponding to times highlighted in red and blue in Fig.~\ref{Fig1}). {Note that these spectra correspond to the same flow, {thus to a run} with the same global parameters, but at different times.} The small-scale PSD of $E_V(k)$ and $E_P(k)$ computed at the local maxima can be up to four orders of magnitude larger than the corresponding PSD for neighboring local minima (see Fig.~\ref{Fig2}). This indicates that vertical drafts excite small-scale turbulent structures, developing in patches within the flow, and powerful enough to modify the spectral distribution of the energy, plausibly with a $k^{-5/3}$ dependence over an inertial range of scales at the times when $K_w$ peaks (see insets in Fig.~\ref{Fig2}). Conversely, the energy spectra computed at the local minima are much lower for $k > 10$, and in fact steeper, plausibly with a $k^{-2}$ scaling at intermediate scales. {The integral of $E_V(k)$ and $E_P(k)$ 
		for $k>30$, act as proxies respectively of the small-scale kinetic and potential energy; they are shown in the inset} of Fig.~\ref{Fig1}. Their correlation with the local maxima and minima of $K_w$ confirms that enhancements of the small-scale PSD are modulated by the extreme vertical drafts.
	
	\begin{figure}[t!]
		\begin{center}
			\includegraphics[width=12cm]{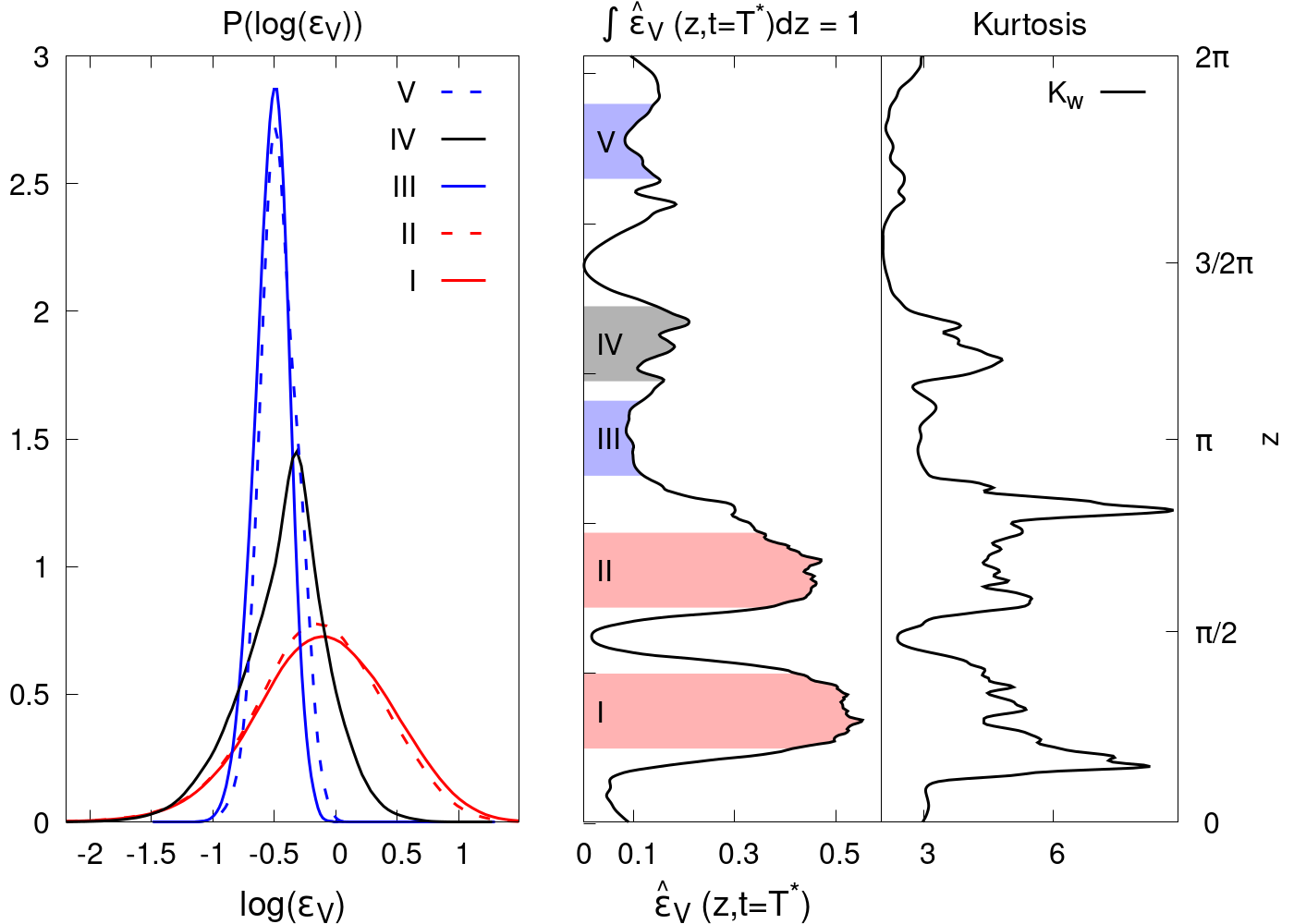}
		\end{center}
		\caption{Left: PDFs of the kinetic energy dissipation computed in the regions indicated by the shaded areas in the middle panel, for run P$5$.
			Middle: profile of the normalized kinetic energy dissipation $\hat{\varepsilon}_V(z,t)$ in individual horizontal planes, at fixed time $T^*\approx 290 \tau_{NL}$ (see Fig. \ref{Fig1}), and as a function of $z$. Right: vertical profile of the by-plane kurtosis $K_w(z,t)$.}
		\label{Fig3}
	\end{figure}
	
	\section{Modulation of kinetic energy dissipation}
	A study of the  statistics of the kinetic and potential energy dissipation rates (respectively $\varepsilon_V=\nu (\partial u_i/\partial x_j)(\partial u_i/\partial x_j)$ and $\varepsilon_P=\kappa |\boldsymbol{\nabla}\zeta|^2$) reveals that the extreme vertical drafts strongly feedback on $\varepsilon_V$ and $\varepsilon_P$, and play a major role in the way the energy is dissipated in stratified turbulence. Fig.~\ref{Fig3} shows the {instantaneous vertical profile of the kurtosis $K_w(z,T^*)$ (i.e., averaged over horizontal planes of constant height)} in run P5, at time $T^*\approx 290\tau_{NL}$, when $K_w(T^*)$ is at a maximum (see Fig.~\ref{Fig1}). The figure {also shows the vertical profile of the kinetic energy dissipation rate $\hat{\varepsilon}_V(z,T^*)$ achieved in horizontal planes, normalized by its value in the entire volume   (so that $\int_0^{2\pi} \hat{\varepsilon}_V(z,T^*) dz = 1$), and the probability density functions (PDFs) of $\hat{\varepsilon}_V$ in regions at different heights.} A comparison between {the profiles of $K_w$ and $\hat{\varepsilon}_V$} reveals some correlation between their peaks (Fig.~\ref{Fig3}, right). The PDFs for regions with strong (I, II), moderate (IV) and weak (III, V) {local} dissipation show that the statistical distribution of the kinetic energy dissipation is modulated by the presence of extreme vertical drafts (Fig.~\ref{Fig3}, left). Indeed $P[\log(\varepsilon_V)]$ is similar between regions with comparable values of $K_w$.
	
	\begin{figure}
		\begin{center}
			\hspace*{-1cm}
			\includegraphics[width=16cm]{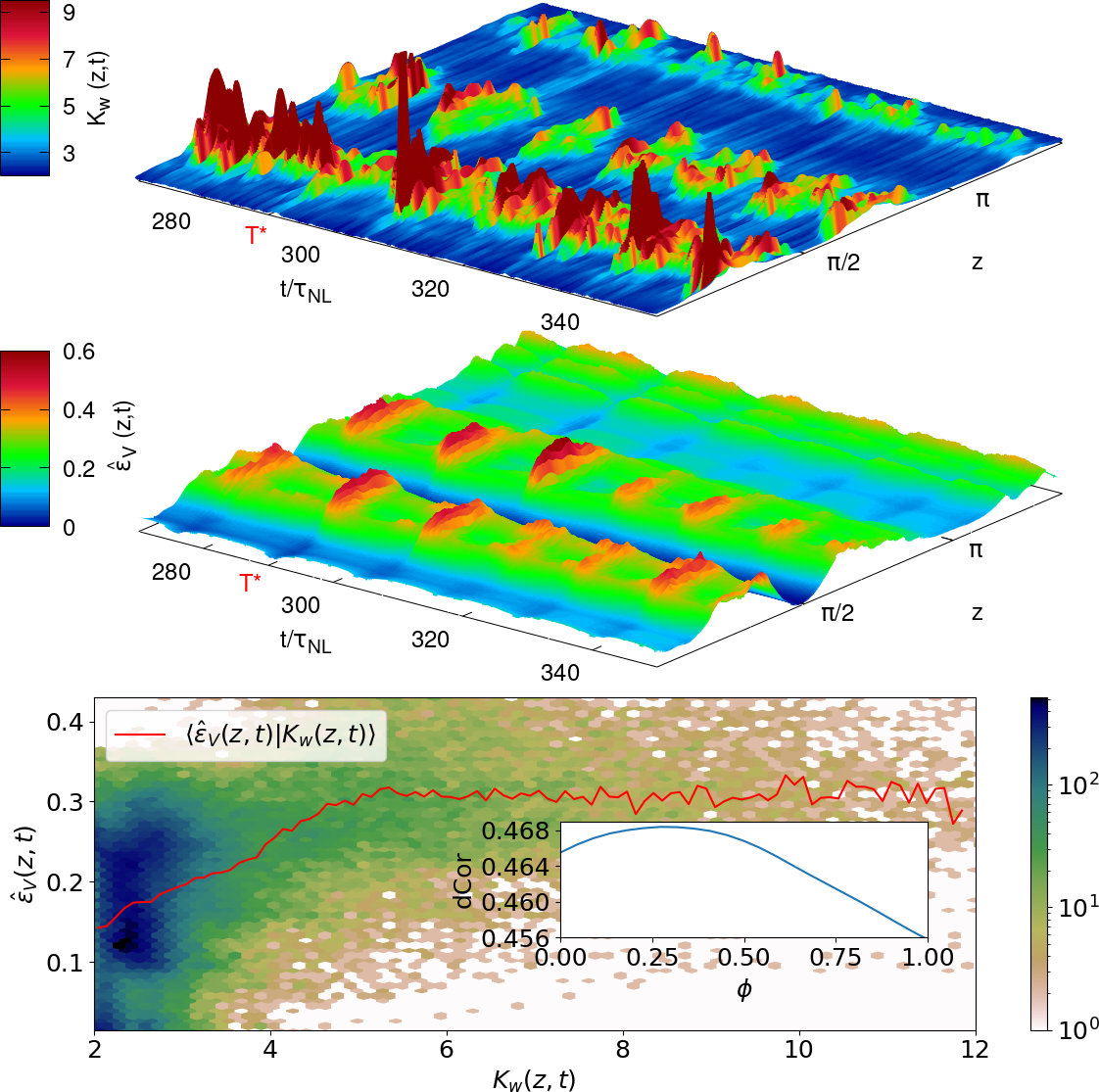}
		\end{center}
		\caption{Temporal evolution in units of $\tau_{NL}$ for run $P5$ of (top) the vertical profile of the by-plane kurtosis $K_w(z,t)$, and (mid-top) the normalized kinetic energy dissipated in horizontal planes $\hat{\varepsilon}_V(z,t)$. 
			The bottom panel gives the joint PDF of $K_w(z,t)$ and $\hat{\varepsilon}_V(z,t)$ for all P5 data, the solid red line indicating averages of $\hat{\varepsilon_V}$ conditioned on $K_w$. 
			Inset: evolution of the distance correlation coefficient {\it dCor} between $K_w(z,t)$ and $\hat{\varepsilon}_V(z,t+\phi)$, as a function of a temporal lag $\phi$ between the two signals. Note that the highest correlation obtains for $\phi \approx \tau_{NL}/3$ (not for $\phi = 0$), thus indicating causation.}
		\label{Fig4}
	\end{figure}
	
	In the top and middle panels, Fig. \ref{Fig4} provides the vertical profiles of the by-plane kurtosis $K_w(z,t)$ and of the normalized kinetic energy dissipation $\hat{\varepsilon}_V(z,t)$, as a function of time (around $T^*$) for run P5. These visualizations emphasize the spatial correlation along the vertical axis $z$ between the emergence of drafts (detected through the amplitude of the kurtosis) and the enhancement of the kinetic energy dissipation, as well as the temporal correlation between these two phenomena. The large peaks of dissipation occur in the same layer of the flow immediately after strong vertical drafts develop. Although the small temporal shift cannot be appreciated from the visualized signals, its existence is demonstrated in the bottom panel of Fig.~\ref{Fig4}, that shows how the (point-wise) values of the quantities rendered in the top-mid panels are maximally correlated for a time delay of $\phi \approx \tau_{NL}/3$, as it results from the analysis of the distance correlation coefficient $dCor_{XY}$, defined next. The latter measures both linear and nonlinear correlations between $X=K_w(z,t)$ and $Y=\hat{\varepsilon}_V(z,t+\phi)$,  for different temporal shifts $\phi$ (shown in the inset in the bottom panel of Fig.~\ref{Fig4}): 
	\begin{equation}
	dCor_{XY} = 
	\frac{\mu_{XY}}{(\mu_{XX}^2\mu_{YY}^2)^{1/4}} ,
	\end{equation}
	where $\mu_{XY},\mu_{XX}$ and $\mu_{YY}$ are correlation and autocorrelation functions defined as in \cite{szekely,edelmann}. It is worth noticing that the overall distance correlation is always rather high, even for $\phi=0$. Finally, the joint PDF of $K_w(z,t)$ and of $\hat{\varepsilon}_V(z,t)$ for all times and heights available for P5 is shown in the bottom panel of Fig.~\ref{Fig4}, together with the conditional average of the dissipation in bins of the kurtosis, $\langle \hat{\varepsilon_V}(z,t)|K_w(z,t) \rangle$ (red line). Note that, locally in space and time, larger values of $K_w$ correspond to larger dissipation rates up to $K_w\approx 6$, while for $K_w > 6$ $\hat{\varepsilon}_V(z,t)$ saturates, these high $K_w$ regions being very efficient at dissipating kinetic energy. The good correlation resulting from the above statistics, together with the evidence that local maxima of $\hat{\varepsilon}_V(z,t)$ are anticipated by peaks in $K_w(z,t)$ {-- the latter occurring $\sim \tau_{NL}/3$ earlier than the former in run P5 --} indicate a causation between the emergence of vertical drafts and the enhancement of the dissipation. Similar results are obtained for other simulations (not shown). Overall, these evidences indicate that the local occurrence of extreme drafts determines the local properties and statistics of strong dissipative events.
	An analysis of the time evolution of the extreme drafts done through renderings of run P5, shows that right after the occurrence of bursts in the vertical velocity, entire horizontal layers of the flow become turbulent displaying strong fluctuations of all the components of the velocity field, both at large- and at small-scales (see movie in the supplemental material \cite{SM}).
	
	\section{Enhanced local dissipation efficiency}
	The efficiency of {the local energy dissipation} can be further characterized by computing the minimal domain volume needed to achieve a given percentage of the global (kinetic and/or potential) energy dissipation at a given time. We therefore evaluate the local kinetic and potential energy dissipation efficiency, respectively $V_{\varepsilon_V}$ and $V_{\varepsilon_P}$, by classifying the temporal outputs of each run in terms of their domain kurtosis $K_w$, and then computing the minimal volume percentage needed to achieve the $50\%$ level of the global energy dissipation.
	
	\begin{figure*}
		\includegraphics[width=13cm]{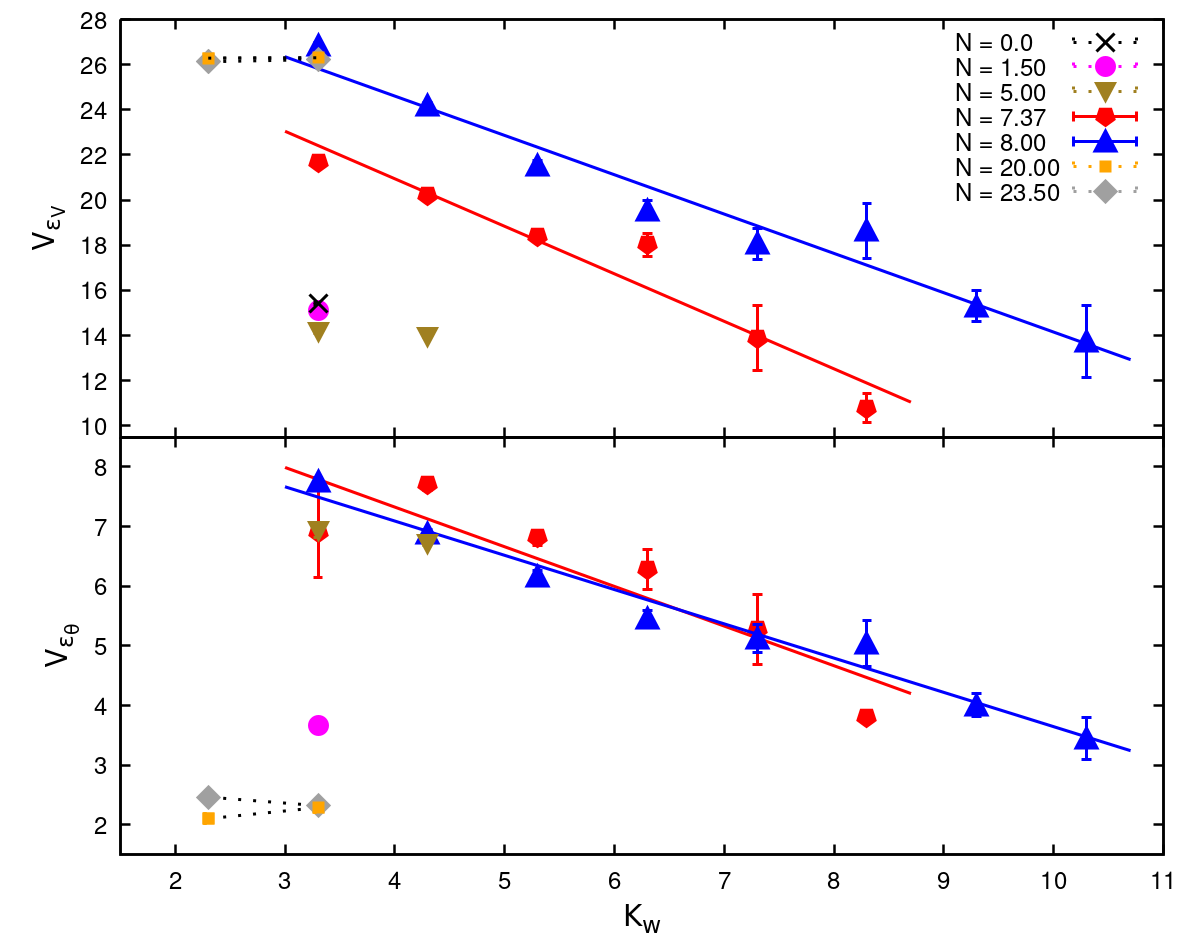}
		\caption{Kinetic and potential energy dissipation efficiency as a function of $K_w$ (for high-Re runs P1--P7), measured in terms of the minimal domain volume percentage $V_{\varepsilon_V}(\%)$ and $V_{\varepsilon_P}(\%)$ needed to achieve $50\%$ of respectively the global kinetic and potential energy dissipation.}
		\label{Fig5}
	\end{figure*}
	
	The outcome of this analysis is shown in Fig.~\ref{Fig5} for the runs with $2400 \leq Re \leq 3800$ (thus within a relatively narrow range of values of the Reynolds number) in order to avoid any Reynolds number dependence of $V_{\varepsilon_V}$ and $V_{\varepsilon_P}$. First, we note that the HIT case has one of the highest kinetic energy dissipation efficiencies: only $\approx 15\%$ of the most dissipative regions within the volume are in fact needed to achieve $50\%$ of the global kinetic energy dissipation (Fig.~\ref{Fig5}, top). Strongly stratified flows are unable to achieve a similar $V_{\varepsilon_V}$ except when they develop extreme vertical drafts, powerful enough for the domain kurtosis to be $K_w \gtrsim 7$, attainable in our study only for Froude numbers within the resonant regime delineated in \cite{feraco2018} (runs P4 and P5), a regime compatible with values found in some regions of the ocean and the atmosphere. Indeed, $V_{\varepsilon_V}$ for these two runs can be respectively as low as $\approx 14\%$ and $\approx 11\%$, smaller in fact than for the HIT case. Thus, not only do the large-scale vertical drafts generate small-scale turbulence, but they are also responsible for the local and efficient enhancement of the kinetic energy dissipation $\varepsilon_V$. These extreme drafts are therefore needed in stratified turbulent flows, when stratification is strong enough ($Fr \lesssim 0.1$), for the energy to be locally dissipated as efficiently as in the HIT case at equivalent Reynolds numbers.
	
	Indeed, without drafts, dissipation efficiency is significantly smaller. The most stratified runs in our study (runs P7 and P8) are unable to develop significant drafts, and they are both characterized by an efficiency $V_{\varepsilon_V}$ $\approx 26\%$, more than twice that of the most dissipative cases of runs P4 and P5. On the opposite limit, when stratification is weak, as for runs P2 and P3, $V_{\varepsilon_V}$ approaches the value of the HIT case (in fact from below) even though $K_w\approx 3$. The local potential energy dissipation efficiency $V_{\varepsilon_P}$ exhibits a behaviour similar to that of $V_{\varepsilon_V}$ (Fig.~\ref{Fig5}, bottom) except for the most stratified runs P7 and P8, that appear to be the most dissipative, although characterized by low kurtosis $K_w$. Moreover, values of $V_{\varepsilon_P}$ are smaller than those of $V_{\varepsilon_V}$, suggesting that stratified flows are more efficient in dissipating potential energy than kinetic energy. This could be related to the well-known stronger small-scale intermittency of (passive) scalars as they easily form frontal structures \cite{pumir_94, sujovolsky_18} (see also \cite{deBruyn_2015}).
	
	\section{Discussion}
	{\subsection{A model of intermittency}} \label{SS:model}
	We showed {that the generation of large-scale intermittent vertical drafts in 
		stratified turbulence~\cite{feraco2018} can 
		lead to recurrent  strong modulations of the flow over duration of up to $\approx 500$ $t/\tau_{NL}$.} These extreme events produce, possibly through instabilities, strong localized {(potential and) kinetic} energy dissipation $\varepsilon_V$ and modulate the overall distribution $P(\varepsilon_V)$, whose shape depends on the region of the flow considered. 
	The presence of vertical drafts is also needed for stably stratified turbulence to achieve a localized dissipation efficiency comparable to that of homogeneous isotropic turbulence. In particular, we showed that, at the peak in $\textrm{Fr}$ of the resonant regime identified in \cite{feraco2018}, roughly 10\% of the domain volume ($V_{\varepsilon_V}$) is sufficient to account for 50\% of the global kinetic energy dissipation.
	
	\begin{figure}[t!]
		\begin{center}
			\includegraphics[trim={0 .1cm 0 1cm},clip,width=15cm]{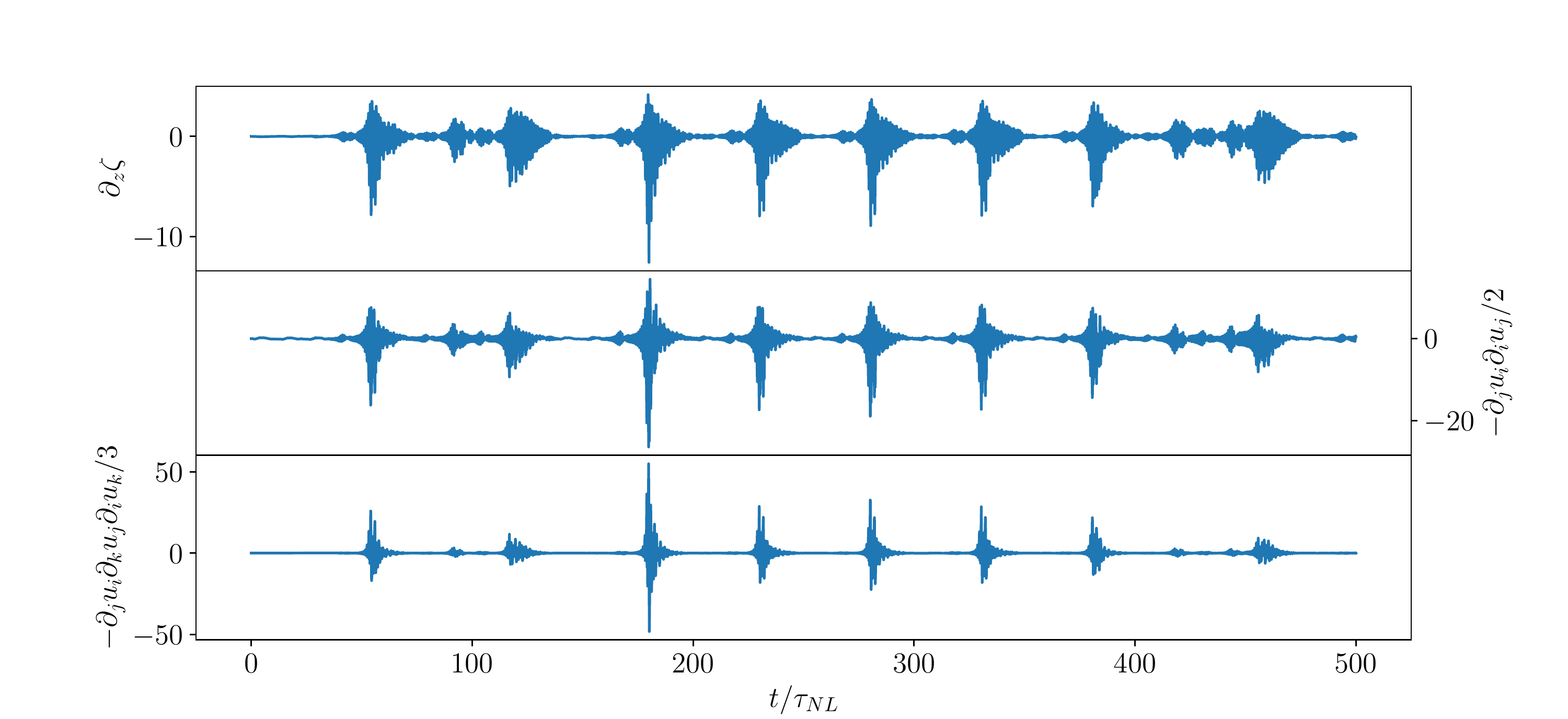}
		\end{center}
		\caption{Time evolution of the vertical gradients of the scaled density fluctuations, $\partial_z \zeta$, and of two different combinations of the velocity field gradients ($Q$ and $R$, two quantities relevant for the evolution of fluid elements and of dissipation in reduced Euler models \cite{Vieillefosse_1984, Meneveau_2011}), in a reduced model of field gradients in stratified flows with parameters compatible with run P5.} \label{Fig6}
	\end{figure}
	
	The kind of intermittency reported here, with slow but strong recurrent modulations, differs from the small-scale intermittency reported in other studies \cite{deBruyn_2015}. The former can be reproduced with a simple model, as we now illustrate. Specifically, we consider a modification of a reduced model for field gradients in stratified turbulence presented in \cite{Sujovolsky_2019, sujovolsky_20},
	\begin{eqnarray}
	\dot{Q} &=& - 3R + NT + f - \gamma Q , \nonumber \\
	\dot{R} &=& 2Q^2/3 + 2NSQ/3 + NR_\zeta - \gamma R , \nonumber \\
	\dot{R_\zeta} &=& 5QT/3 + 3RS - 4NST/3 - NQA - NR - \gamma R_\zeta , \nonumber \\
	\dot{B} &=& 2QA/3 + 2R - NAS/3 - NT - \gamma B , \nonumber \\
	\dot{T} &=& - 2R_\zeta - 2SQ/3 + NB - 2NS^2/3 - \gamma T , \nonumber \\
	\dot{A} &=& - B - 2Q/3 - 2NS/3 - \gamma A , \nonumber \\
	\dot{S} &=& NA - T - \gamma S , \nonumber
	\end{eqnarray}
	where $N$ is the Brunt-V\"ais\"al\"a frequency as before,
	$Q=(\partial_j u_i)(\partial_i u_j)/2$, $R = (\partial_j u_i)(\partial_k u_j)(\partial_i u_k)/3$, $R_\zeta = (\partial_i \zeta)(\partial_j u_i) (\partial_z u_j)$, $B = (\partial_i u_z)(\partial_z u_i)$, $T= (\partial_i \zeta)(\partial_z u_i)$, $A = \partial_z u_z$, and $S= \partial_z \zeta$ (i.e., all combinations of field gradients). Viscous damping (controlled by $\gamma$) and forcing were added phenomenologically. These equations (for $f=\gamma=0$) can be derived from Eqs.~(\ref{bequationV}) and (\ref{bequationT}) following fluid trajectories  \cite{sujovolsky_20}. We integrate these equations with $f$ given by a superposition of harmonic oscillations with frequencies centered around the Brunt-V\"ais\"al\"a frequency $N$ and with very small amplitude ($=0.08$), and with $\gamma = \nu/L_\textrm{Oz}^2$, where $L_\textrm{Oz}$ is the Ozmidov length. 
	{This length is defined as $L_\textrm{Oz}=\sqrt{\varepsilon_V/N^3}$; it can be viewed as partitioning the flow between larger scales governed by quasi geostrophic dynamics which progressively gives way, at smaller scales, to  strong stratified turbulence.} 
	The values of $N$, $\nu$, and $L_\textrm{Oz}$ where chosen as in run P5. The result is shown in Fig.~\ref{Fig6}. Note the system is bursty with a behavior reminiscent of 
	so-called on-off intermittency \cite{pomeau_80, ott_94, saha_18}:
	it displays long periods of very small oscillations, followed by non-regular (but repetitive) bursts reminding those observed in Fig.~\ref{Fig1}, which are separated by times much larger than the typical time scales in the system. Also, both $\partial_z \theta$ as well as $\partial_z w$ (not shown) have bursts, as well as $Q$ and $R$, two quantities relevant in many reduced Euler models \cite{Vieillefosse_1984, Meneveau_2011} to describe vortex stretching and dissipation. The bursts, amplifying the forcing by orders of magnitude, take place as the system evolves between two slow manifolds \cite{sujovolsky_20}, and the overall behavior is compatible with a stochastic resonance \cite{benzi_82}.
	
	{\subsection{Conclusion}}
	The observation of the strong spatial localization of the dissipation in our DNSs is reminiscent of results presented in \cite{pearson} using global oceanic simulations. Indeed, Pearson and Fox-Kemper have shown that macroscopic features of the PDF of $\varepsilon_V$ in their simulations (i.e., the mean and standard deviation) depend on the depth and the sub-domains considered, concluding that most of the dissipation at oceanic mesoscales occurs in a small number of high-dissipation locations corresponding to a fraction of the ocean volume. It is worth noticing, however, that the model used in \cite{pearson} includes subgrid modelling of the dissipation, plus the effect of topography as well as other relevant effects for global oceanic modelling. By showing that large-scale intermittent structures emerging in the bulk of stratified flows are associated with enhanced kinetic energy dissipation, our work indicates that the underlying mechanism associated with the development of regions of extreme dissipation may be a fundamental property of turbulence in the presence of stable stratification, as also suggested by the simple model that reproduces the recurrent bursts 
	(see \S \ref{SS:model}). 
	Our results shed also light on the link between intermittency and dissipation recently emphasized in \cite{turiel}.
	
	{The way energy is dissipated in geophysical flows remains an important open problem; our study indicates that in a certain region of parameter space,} vertical drafts and the associated steepening of gravity waves can lead to enhanced local-in-time-and-space dissipation, ultimately leading to an inadequacy of the description of the system in terms of solely weakly interacting waves, even when the global Froude number is small. The state of marginal instability and its relationship with the efficiency of energy dissipation and mixing, also in the context of ocean dynamics, has been analyzed recently \cite{smyth_20}. Using a model, it was shown that it may be governed by regions of the flow close to a margin of instability for the Richardson number, consistently with results obtained from other reduced models \cite{sujovolsky_20}.
	{Finally, we mention important extensions of this work. With its focus on intermittency, our study is necessarily of a statistical nature, as quantities such as the kurtosis are only defined by averages. As a result, the present analysis could show that stratified turbulence can dissipate energy as efficiently as homogeneous and isotropic turbulence for some values of the Froude number, but it cannot pinpoint individual structures responsible for the dissipation, or characterize the dynamics of such structures. From the point of view of out-of-equilibrium statistical mechanics, the origin of these events can be understood, with the help of simple dynamical models involving a nonlinear resonant-like amplification of waves by eddies \cite{rorai2014, feraco2018, feraco2021}, as a self-organized critical process in which the strong events lead to a cascade of smaller scale extreme events \cite{Smyth_2019, pouquet_19p}. Another way to analyze the dynamics is by postulating the existence of two slow manifolds in the dynamics (one associated with waves, the other with the overturning instability) with fluid elements evolving fast from one available state to the other \cite{sujovolsky_20}, as in the model briefly discussed in this section. A study of the fluid dynamics of the evolution of individual structures, in the spirit of traditional studies of stably stratified turbulence as, e.g., in \cite{Kimura_1996, Billant_2000}, or in \cite{Winters_1995} considering the role of conserved quantities, is left for the future.}
	
	\begin{acknowledgments}
		R. Marino acknowledges support from the project ``EVENTFUL" (ANR-20-CE30-0011), funded by the French ``Agence Nationale de la Recherche" - ANR through the program AAPG-2020. A. Pouquet is thankful to LASP and in particular to Bob Ergun. 
	\end{acknowledgments}
	
	
	\bibliography{ms}

\begin{thebibliography}{71}%
\makeatletter
\providecommand \@ifxundefined [1]{%
 \@ifx{#1\undefined}
}%
\providecommand \@ifnum [1]{%
 \ifnum #1\expandafter \@firstoftwo
 \else \expandafter \@secondoftwo
 \fi
}%
\providecommand \@ifx [1]{%
 \ifx #1\expandafter \@firstoftwo
 \else \expandafter \@secondoftwo
 \fi
}%
\providecommand \natexlab [1]{#1}%
\providecommand \enquote  [1]{``#1''}%
\providecommand \bibnamefont  [1]{#1}%
\providecommand \bibfnamefont [1]{#1}%
\providecommand \citenamefont [1]{#1}%
\providecommand \href@noop [0]{\@secondoftwo}%
\providecommand \href [0]{\begingroup \@sanitize@url \@href}%
\providecommand \@href[1]{\@@startlink{#1}\@@href}%
\providecommand \@@href[1]{\endgroup#1\@@endlink}%
\providecommand \@sanitize@url [0]{\catcode `\\12\catcode `\$12\catcode
  `\&12\catcode `\#12\catcode `\^12\catcode `\_12\catcode `\%12\relax}%
\providecommand \@@startlink[1]{}%
\providecommand \@@endlink[0]{}%
\providecommand \url  [0]{\begingroup\@sanitize@url \@url }%
\providecommand \@url [1]{\endgroup\@href {#1}{\urlprefix }}%
\providecommand \urlprefix  [0]{URL }%
\providecommand \Eprint [0]{\href }%
\providecommand \doibase [0]{http://dx.doi.org/}%
\providecommand \selectlanguage [0]{\@gobble}%
\providecommand \bibinfo  [0]{\@secondoftwo}%
\providecommand \bibfield  [0]{\@secondoftwo}%
\providecommand \translation [1]{[#1]}%
\providecommand \BibitemOpen [0]{}%
\providecommand \bibitemStop [0]{}%
\providecommand \bibitemNoStop [0]{.\EOS\space}%
\providecommand \EOS [0]{\spacefactor3000\relax}%
\providecommand \BibitemShut  [1]{\csname bibitem#1\endcsname}%
\let\auto@bib@innerbib\@empty
\bibitem [{\citenamefont {Marcus}(1993)}]{marcus_93}%
  \BibitemOpen
  \bibfield  {author} {\bibinfo {author} {\bibfnamefont {P.~S.}\ \bibnamefont
  {Marcus}},\ }\bibfield  {title} {\enquote {\bibinfo {title} {Jupiter's great
  red spot and other vortices},}\ }\href@noop {} {\bibfield  {journal}
  {\bibinfo  {journal} {Annu. Rev. Astron. Astrophys.}\ }\textbf {\bibinfo
  {volume} {31}},\ \bibinfo {pages} {523--573} (\bibinfo {year}
  {1993})}\BibitemShut {NoStop}%
\bibitem [{\citenamefont {Emanuel}(2005)}]{emanuel_05}%
  \BibitemOpen
  \bibfield  {author} {\bibinfo {author} {\bibfnamefont {K.}~\bibnamefont
  {Emanuel}},\ }\bibfield  {title} {\enquote {\bibinfo {title} {Increasing
  destructiveness of tropical cyclones over the past 30 years},}\ }\href@noop
  {} {\bibfield  {journal} {\bibinfo  {journal} {Nature}\ }\textbf {\bibinfo
  {volume} {436}},\ \bibinfo {pages} {686--688} (\bibinfo {year}
  {2005})}\BibitemShut {NoStop}%
\bibitem [{\citenamefont {Scott}\ and\ \citenamefont {Wang}(2005)}]{scott_05}%
  \BibitemOpen
  \bibfield  {author} {\bibinfo {author} {\bibfnamefont {R.B.}\ \bibnamefont
  {Scott}}\ and\ \bibinfo {author} {\bibfnamefont {F.}~\bibnamefont {Wang}},\
  }\bibfield  {title} {\enquote {\bibinfo {title} {Direct evidence of an
  oceanic inverse kinetic energy cascade from satellite altimetry},}\
  }\href@noop {} {\bibfield  {journal} {\bibinfo  {journal} {J. Phys. Oceano.}\
  }\textbf {\bibinfo {volume} {35}},\ \bibinfo {pages} {1650--1666} (\bibinfo
  {year} {2005})}\BibitemShut {NoStop}%
\bibitem [{\citenamefont {Klymak}\ \emph {et~al.}(2008)\citenamefont {Klymak},
  \citenamefont {Pinkel},\ and\ \citenamefont {Rainville}}]{klymak_08}%
  \BibitemOpen
  \bibfield  {author} {\bibinfo {author} {\bibfnamefont {J.M.}\ \bibnamefont
  {Klymak}}, \bibinfo {author} {\bibfnamefont {R.}~\bibnamefont {Pinkel}}, \
  and\ \bibinfo {author} {\bibfnamefont {L.}~\bibnamefont {Rainville}},\
  }\bibfield  {title} {\enquote {\bibinfo {title} {Direct breaking of the
  internal tide near topography: {K}aena {R}idge, {H}awaii},}\ }\href@noop {}
  {\bibfield  {journal} {\bibinfo  {journal} {J. Phys. Oceano.}\ }\textbf
  {\bibinfo {volume} {38}},\ \bibinfo {pages} {380--399} (\bibinfo {year}
  {2008})}\BibitemShut {NoStop}%
\bibitem [{\citenamefont {Klein}\ \emph {et~al.}(2008)\citenamefont {Klein},
  \citenamefont {Hua}, \citenamefont {Lapeyre}, \citenamefont {Capet},
  \citenamefont {Gentil},\ and\ \citenamefont {Sasaki}}]{klein_08}%
  \BibitemOpen
  \bibfield  {author} {\bibinfo {author} {\bibfnamefont {P.}~\bibnamefont
  {Klein}}, \bibinfo {author} {\bibfnamefont {B.L.}\ \bibnamefont {Hua}},
  \bibinfo {author} {\bibfnamefont {G.}~\bibnamefont {Lapeyre}}, \bibinfo
  {author} {\bibfnamefont {X.}~\bibnamefont {Capet}}, \bibinfo {author}
  {\bibfnamefont {S.~Le}\ \bibnamefont {Gentil}}, \ and\ \bibinfo {author}
  {\bibfnamefont {H.}~\bibnamefont {Sasaki}},\ }\bibfield  {title} {\enquote
  {\bibinfo {title} {Upper ocean turbulence from high-resolution 3{D}
  simulations},}\ }\href@noop {} {\bibfield  {journal} {\bibinfo  {journal} {J.
  Phys. Oceano.}\ }\textbf {\bibinfo {volume} {38}},\ \bibinfo {pages}
  {1748--1763} (\bibinfo {year} {2008})}\BibitemShut {NoStop}%
\bibitem [{\citenamefont {Marino}\ \emph {et~al.}(2013)\citenamefont {Marino},
  \citenamefont {Mininni}, \citenamefont {Rosenberg},\ and\ \citenamefont
  {Pouquet}}]{marino_13i}%
  \BibitemOpen
  \bibfield  {author} {\bibinfo {author} {\bibfnamefont {R.}~\bibnamefont
  {Marino}}, \bibinfo {author} {\bibfnamefont {P.D.}\ \bibnamefont {Mininni}},
  \bibinfo {author} {\bibfnamefont {D.}~\bibnamefont {Rosenberg}}, \ and\
  \bibinfo {author} {\bibfnamefont {A.}~\bibnamefont {Pouquet}},\ }\bibfield
  {title} {\enquote {\bibinfo {title} {Inverse cascades in rotating stratified
  turbulence: Fast growth of large scales},}\ }\href@noop {} {\bibfield
  {journal} {\bibinfo  {journal} {Eur. Phys. Lett.}\ }\textbf {\bibinfo
  {volume} {102}},\ \bibinfo {pages} {44006} (\bibinfo {year}
  {2013})}\BibitemShut {NoStop}%
\bibitem [{\citenamefont {Pouquet}\ and\ \citenamefont
  {Marino}(2013)}]{pouquet_13}%
  \BibitemOpen
  \bibfield  {author} {\bibinfo {author} {\bibfnamefont {A.}~\bibnamefont
  {Pouquet}}\ and\ \bibinfo {author} {\bibfnamefont {R.}~\bibnamefont
  {Marino}},\ }\bibfield  {title} {\enquote {\bibinfo {title} {Geophysical
  turbulence and the duality of the energy flow across scales},}\ }\href@noop
  {} {\bibfield  {journal} {\bibinfo  {journal} {Phys. Rev. Lett.}\ }\textbf
  {\bibinfo {volume} {234501}},\ \bibinfo {pages} {111} (\bibinfo {year}
  {2013})}\BibitemShut {NoStop}%
\bibitem [{\citenamefont {Marino}\ \emph {et~al.}(2014)\citenamefont {Marino},
  \citenamefont {Mininni}, \citenamefont {Rosenberg},\ and\ \citenamefont
  {Pouquet}}]{marino_14}%
  \BibitemOpen
  \bibfield  {author} {\bibinfo {author} {\bibfnamefont {R.}~\bibnamefont
  {Marino}}, \bibinfo {author} {\bibfnamefont {P.D.}\ \bibnamefont {Mininni}},
  \bibinfo {author} {\bibfnamefont {D.}~\bibnamefont {Rosenberg}}, \ and\
  \bibinfo {author} {\bibfnamefont {A.}~\bibnamefont {Pouquet}},\ }\bibfield
  {title} {\enquote {\bibinfo {title} {Large-scale anisotropy in stably
  stratified rotating flows},}\ }\href@noop {} {\bibfield  {journal} {\bibinfo
  {journal} {Phys. Rev. E}\ }\textbf {\bibinfo {volume} {90}},\ \bibinfo
  {pages} {023018} (\bibinfo {year} {2014})}\BibitemShut {NoStop}%
\bibitem [{\citenamefont {Marino}\ \emph
  {et~al.}(2015{\natexlab{a}})\citenamefont {Marino}, \citenamefont {Pouquet},\
  and\ \citenamefont {Rosenberg}}]{marino_15p}%
  \BibitemOpen
  \bibfield  {author} {\bibinfo {author} {\bibfnamefont {R.}~\bibnamefont
  {Marino}}, \bibinfo {author} {\bibfnamefont {A.}~\bibnamefont {Pouquet}}, \
  and\ \bibinfo {author} {\bibfnamefont {D.}~\bibnamefont {Rosenberg}},\
  }\bibfield  {title} {\enquote {\bibinfo {title} {Resolving the paradox of
  oceanic large-scale balance and small-scale mixing},}\ }\href@noop {}
  {\bibfield  {journal} {\bibinfo  {journal} {Phys. Rev. Lett.}\ }\textbf
  {\bibinfo {volume} {114}},\ \bibinfo {pages} {114504} (\bibinfo {year}
  {2015}{\natexlab{a}})}\BibitemShut {NoStop}%
\bibitem [{\citenamefont {Pouquet}\ \emph {et~al.}(2017)\citenamefont
  {Pouquet}, \citenamefont {Marino}, \citenamefont {Mininni},\ and\
  \citenamefont {Rosenberg}}]{pouquet_17}%
  \BibitemOpen
  \bibfield  {author} {\bibinfo {author} {\bibfnamefont {A.}~\bibnamefont
  {Pouquet}}, \bibinfo {author} {\bibfnamefont {R.}~\bibnamefont {Marino}},
  \bibinfo {author} {\bibfnamefont {P.D.}\ \bibnamefont {Mininni}}, \ and\
  \bibinfo {author} {\bibfnamefont {D.}~\bibnamefont {Rosenberg}},\ }\bibfield
  {title} {\enquote {\bibinfo {title} {Dual constant-flux energy cascades to
  both large scales and small scales},}\ }\href@noop {} {\bibfield  {journal}
  {\bibinfo  {journal} {Phys. Fluids}\ }\textbf {\bibinfo {volume} {29}},\
  \bibinfo {pages} {111108} (\bibinfo {year} {2017})}\BibitemShut {NoStop}%
\bibitem [{\citenamefont {Pouquet}\ \emph
  {et~al.}(2019{\natexlab{a}})\citenamefont {Pouquet}, \citenamefont
  {Rosenberg}, \citenamefont {Stawarz},\ and\ \citenamefont
  {Marino}}]{pouquet_19}%
  \BibitemOpen
  \bibfield  {author} {\bibinfo {author} {\bibfnamefont {A.}~\bibnamefont
  {Pouquet}}, \bibinfo {author} {\bibfnamefont {D.}~\bibnamefont {Rosenberg}},
  \bibinfo {author} {\bibfnamefont {J.E.}\ \bibnamefont {Stawarz}}, \ and\
  \bibinfo {author} {\bibfnamefont {R.}~\bibnamefont {Marino}},\ }\bibfield
  {title} {\enquote {\bibinfo {title} {Helicity dynamics, inverse, and
  bidirectional cascades in fluid and magnetohydrodynamic turbulence: A brief
  review},}\ }\href@noop {} {\bibfield  {journal} {\bibinfo  {journal} {Earth
  and Space Science}\ }\textbf {\bibinfo {volume} {6}},\ \bibinfo {pages}
  {351--369} (\bibinfo {year} {2019}{\natexlab{a}})}\BibitemShut {NoStop}%
\bibitem [{\citenamefont {Pouquet}\ \emph
  {et~al.}(2019{\natexlab{b}})\citenamefont {Pouquet}, \citenamefont
  {Rosenberg}, \citenamefont {Stawarz},\ and\ \citenamefont
  {Marino}}]{pouquet_19e}%
  \BibitemOpen
  \bibfield  {author} {\bibinfo {author} {\bibfnamefont {A.}~\bibnamefont
  {Pouquet}}, \bibinfo {author} {\bibfnamefont {D.}~\bibnamefont {Rosenberg}},
  \bibinfo {author} {\bibfnamefont {J.}~\bibnamefont {Stawarz}}, \ and\
  \bibinfo {author} {\bibfnamefont {R.}~\bibnamefont {Marino}},\ }\bibfield
  {title} {\enquote {\bibinfo {title} {Helicity dynamics, inverse, and
  bidirectional cascades in fluid and magnetohydrodynamic turbulence: A brief
  review},}\ }\href@noop {} {\bibfield  {journal} {\bibinfo  {journal} {Earth
  Space Sci.}\ }\textbf {\bibinfo {volume} {6}},\ \bibinfo {pages} {1--19}
  (\bibinfo {year} {2019}{\natexlab{b}})}\BibitemShut {NoStop}%
\bibitem [{\citenamefont {Xie}\ and\ \citenamefont {B{\"u}hler}(2019)}]{xie1}%
  \BibitemOpen
  \bibfield  {author} {\bibinfo {author} {\bibfnamefont {J.H.}\ \bibnamefont
  {Xie}}\ and\ \bibinfo {author} {\bibfnamefont {O.}~\bibnamefont
  {B{\"u}hler}},\ }\bibfield  {title} {\enquote {\bibinfo {title}
  {Two-dimensional isotropic inertia--gravity wave turbulence},}\ }\href@noop
  {} {\bibfield  {journal} {\bibinfo  {journal} {J. Fluid Mech.}\ }\textbf
  {\bibinfo {volume} {872}},\ \bibinfo {pages} {752--783} (\bibinfo {year}
  {2019})}\BibitemShut {NoStop}%
\bibitem [{\citenamefont {Xie}\ and\ \citenamefont {B{\"u}hler}(2018)}]{xie2}%
  \BibitemOpen
  \bibfield  {author} {\bibinfo {author} {\bibfnamefont {J.H.}\ \bibnamefont
  {Xie}}\ and\ \bibinfo {author} {\bibfnamefont {O.}~\bibnamefont
  {B{\"u}hler}},\ }\bibfield  {title} {\enquote {\bibinfo {title} {Exact
  third-order structure functions for two-dimensional turbulence},}\
  }\href@noop {} {\bibfield  {journal} {\bibinfo  {journal} {J. Fluid Mech.}\
  }\textbf {\bibinfo {volume} {851}},\ \bibinfo {pages} {672--686} (\bibinfo
  {year} {2018})}\BibitemShut {NoStop}%
\bibitem [{\citenamefont {Marino}\ \emph
  {et~al.}(2015{\natexlab{b}})\citenamefont {Marino}, \citenamefont
  {Rosenberg}, \citenamefont {Herbert},\ and\ \citenamefont
  {Pouquet}}]{marino_15w}%
  \BibitemOpen
  \bibfield  {author} {\bibinfo {author} {\bibfnamefont {R.}~\bibnamefont
  {Marino}}, \bibinfo {author} {\bibfnamefont {D.}~\bibnamefont {Rosenberg}},
  \bibinfo {author} {\bibfnamefont {C.}~\bibnamefont {Herbert}}, \ and\
  \bibinfo {author} {\bibfnamefont {A.}~\bibnamefont {Pouquet}},\ }\bibfield
  {title} {\enquote {\bibinfo {title} {Interplay of waves and eddies in
  rotating stratified turbulence and the link with kinetic-potential energy
  partition},}\ }\href@noop {} {\bibfield  {journal} {\bibinfo  {journal} {Eur.
  Phys. Lett.}\ }\textbf {\bibinfo {volume} {112}},\ \bibinfo {pages} {49001}
  (\bibinfo {year} {2015}{\natexlab{b}})}\BibitemShut {NoStop}%
\bibitem [{\citenamefont {Pouquet}\ \emph {et~al.}(2018)\citenamefont
  {Pouquet}, \citenamefont {Rosenberg}, \citenamefont {Marino},\ and\
  \citenamefont {Herbert}}]{pouquet_18j}%
  \BibitemOpen
  \bibfield  {author} {\bibinfo {author} {\bibfnamefont {A.}~\bibnamefont
  {Pouquet}}, \bibinfo {author} {\bibfnamefont {D.}~\bibnamefont {Rosenberg}},
  \bibinfo {author} {\bibfnamefont {R.}~\bibnamefont {Marino}}, \ and\ \bibinfo
  {author} {\bibfnamefont {C.}~\bibnamefont {Herbert}},\ }\bibfield  {title}
  {\enquote {\bibinfo {title} {Scaling laws for mixing and dissipation in
  unforced rotating stratified turbulence},}\ }\href@noop {} {\bibfield
  {journal} {\bibinfo  {journal} {J. Fluid Mech.}\ }\textbf {\bibinfo {volume}
  {844}},\ \bibinfo {pages} {519--545} (\bibinfo {year} {2018})}\BibitemShut
  {NoStop}%
\bibitem [{\citenamefont {Wang}\ and\ \citenamefont
  {B{\"u}hler}(2020)}]{buhler}%
  \BibitemOpen
  \bibfield  {author} {\bibinfo {author} {\bibfnamefont {H.}~\bibnamefont
  {Wang}}\ and\ \bibinfo {author} {\bibfnamefont {O.}~\bibnamefont
  {B{\"u}hler}},\ }\bibfield  {title} {\enquote {\bibinfo {title} {Ageostrophic
  corrections for power spectra and wave--vortex decomposition},}\ }\href@noop
  {} {\bibfield  {journal} {\bibinfo  {journal} {J. Fluid Mech.}\ }\textbf
  {\bibinfo {volume} {882}},\ \bibinfo {pages} {A16} (\bibinfo {year}
  {2020})}\BibitemShut {NoStop}%
\bibitem [{\citenamefont {Herbert}\ \emph {et~al.}(2016)\citenamefont
  {Herbert}, \citenamefont {Marino}, \citenamefont {Rosenberg},\ and\
  \citenamefont {Pouquet}}]{herbert_16}%
  \BibitemOpen
  \bibfield  {author} {\bibinfo {author} {\bibfnamefont {C.}~\bibnamefont
  {Herbert}}, \bibinfo {author} {\bibfnamefont {R.}~\bibnamefont {Marino}},
  \bibinfo {author} {\bibfnamefont {D.}~\bibnamefont {Rosenberg}}, \ and\
  \bibinfo {author} {\bibfnamefont {A.}~\bibnamefont {Pouquet}},\ }\bibfield
  {title} {\enquote {\bibinfo {title} {Waves and vortices in the inverse
  cascade regime of stratified turbulence with or without rotation},}\
  }\href@noop {} {\bibfield  {journal} {\bibinfo  {journal} {Journal of Fluid
  Mechanics}\ }\textbf {\bibinfo {volume} {806}},\ \bibinfo {pages} {165--204}
  (\bibinfo {year} {2016})}\BibitemShut {NoStop}%
\bibitem [{\citenamefont {Lenschow}\ \emph {et~al.}(2012)\citenamefont
  {Lenschow}, \citenamefont {Lothon}, \citenamefont {Mayor}, \citenamefont
  {Sullivan},\ and\ \citenamefont {Canut}}]{lenschow_12}%
  \BibitemOpen
  \bibfield  {author} {\bibinfo {author} {\bibfnamefont {D.~H.}\ \bibnamefont
  {Lenschow}}, \bibinfo {author} {\bibfnamefont {M.}~\bibnamefont {Lothon}},
  \bibinfo {author} {\bibfnamefont {S.~D.}\ \bibnamefont {Mayor}}, \bibinfo
  {author} {\bibfnamefont {P.~P.}\ \bibnamefont {Sullivan}}, \ and\ \bibinfo
  {author} {\bibfnamefont {G.}~\bibnamefont {Canut}},\ }\bibfield  {title}
  {\enquote {\bibinfo {title} {A comparison of higher-order vertical velocity
  moments in the convective boundary layer from {L}idar with in situ
  measurements and {L}arge-{E}ddy {S}imulation},}\ }\href@noop {} {\bibfield
  {journal} {\bibinfo  {journal} {Bound. Lay. Met.}\ }\textbf {\bibinfo
  {volume} {143}},\ \bibinfo {pages} {107--123} (\bibinfo {year}
  {2012})}\BibitemShut {NoStop}%
\bibitem [{\citenamefont {Mahrt}\ and\ \citenamefont {Gamage}(1987)}]{mahrt1}%
  \BibitemOpen
  \bibfield  {author} {\bibinfo {author} {\bibfnamefont {L.}~\bibnamefont
  {Mahrt}}\ and\ \bibinfo {author} {\bibfnamefont {N.}~\bibnamefont {Gamage}},\
  }\bibfield  {title} {\enquote {\bibinfo {title} {Observations of turbulence
  in stratified flow},}\ }\href@noop {} {\bibfield  {journal} {\bibinfo
  {journal} {J. Atmos. Sci.}\ }\textbf {\bibinfo {volume} {44}},\ \bibinfo
  {pages} {1106--1122} (\bibinfo {year} {1987})}\BibitemShut {NoStop}%
\bibitem [{\citenamefont {Mahrt}(1989)}]{mahrt2}%
  \BibitemOpen
  \bibfield  {author} {\bibinfo {author} {\bibfnamefont {L.}~\bibnamefont
  {Mahrt}},\ }\bibfield  {title} {\enquote {\bibinfo {title} {Intermittency of
  atmospheric turbulence},}\ }\href@noop {} {\bibfield  {journal} {\bibinfo
  {journal} {J. Atmos. Sci.}\ }\textbf {\bibinfo {volume} {46}},\ \bibinfo
  {pages} {79 -- 95} (\bibinfo {year} {1989})}\BibitemShut {NoStop}%
\bibitem [{\citenamefont {Liu}(2007)}]{liu_07}%
  \BibitemOpen
  \bibfield  {author} {\bibinfo {author} {\bibfnamefont {H.L.}\ \bibnamefont
  {Liu}},\ }\bibfield  {title} {\enquote {\bibinfo {title} {On the large wind
  shear and fast meridional transport above the mesopause},}\ }\href@noop {}
  {\bibfield  {journal} {\bibinfo  {journal} {Geophys. Res. Lett.}\ }\textbf
  {\bibinfo {volume} {34}},\ \bibinfo {pages} {L08815} (\bibinfo {year}
  {2007})}\BibitemShut {NoStop}%
\bibitem [{\citenamefont {Chau}\ \emph {et~al.}(2021)\citenamefont {Chau},
  \citenamefont {Marino}, \citenamefont {Feraco}, \citenamefont {Cordero},
  \citenamefont {Baumgarten}, \citenamefont {Luebken}, \citenamefont {Hocking},
  \citenamefont {Schult}, \citenamefont {Renkwitz},\ and\ \citenamefont
  {Latteck}}]{chau_21}%
  \BibitemOpen
  \bibfield  {author} {\bibinfo {author} {\bibfnamefont {J.L.}\ \bibnamefont
  {Chau}}, \bibinfo {author} {\bibfnamefont {R.}~\bibnamefont {Marino}},
  \bibinfo {author} {\bibfnamefont {F.}~\bibnamefont {Feraco}}, \bibinfo
  {author} {\bibfnamefont {J.M.~Urco}\ \bibnamefont {Cordero}}, \bibinfo
  {author} {\bibfnamefont {G.}~\bibnamefont {Baumgarten}}, \bibinfo {author}
  {\bibfnamefont {F-J.}\ \bibnamefont {Luebken}}, \bibinfo {author}
  {\bibfnamefont {W.K.}\ \bibnamefont {Hocking}}, \bibinfo {author}
  {\bibfnamefont {C.}~\bibnamefont {Schult}}, \bibinfo {author} {\bibfnamefont
  {T.}~\bibnamefont {Renkwitz}}, \ and\ \bibinfo {author} {\bibfnamefont
  {R.}~\bibnamefont {Latteck}},\ }\bibfield  {title} {\enquote {\bibinfo
  {title} {Radar observation of extreme vertical drafts in the polar summer
  mesosphere},}\ }\href@noop {} {\bibfield  {journal} {\bibinfo  {journal}
  {Geophys. Res. Lett.}\ }\textbf {\bibinfo {volume} {48}},\ \bibinfo {pages}
  {e2021GL094918} (\bibinfo {year} {2021})}\BibitemShut {NoStop}%
\bibitem [{\citenamefont {D'Asaro}\ \emph {et~al.}(2007)\citenamefont
  {D'Asaro}, \citenamefont {Lien},\ and\ \citenamefont {Henyey}}]{dasaro}%
  \BibitemOpen
  \bibfield  {author} {\bibinfo {author} {\bibfnamefont {E.}~\bibnamefont
  {D'Asaro}}, \bibinfo {author} {\bibfnamefont {R-C.}\ \bibnamefont {Lien}}, \
  and\ \bibinfo {author} {\bibfnamefont {F.}~\bibnamefont {Henyey}},\
  }\bibfield  {title} {\enquote {\bibinfo {title} {High-frequency internal
  waves on the oregon continental shelf},}\ }\href@noop {} {\bibfield
  {journal} {\bibinfo  {journal} {J. Phys. Oceanogr.}\ }\textbf {\bibinfo
  {volume} {37}},\ \bibinfo {pages} {1956--1967} (\bibinfo {year}
  {2007})}\BibitemShut {NoStop}%
\bibitem [{\citenamefont {van Haren}\ and\ \citenamefont
  {Gostiaux}(2016)}]{vanharen_16j}%
  \BibitemOpen
  \bibfield  {author} {\bibinfo {author} {\bibfnamefont {H.}~\bibnamefont {van
  Haren}}\ and\ \bibinfo {author} {\bibfnamefont {L.}~\bibnamefont
  {Gostiaux}},\ }\bibfield  {title} {\enquote {\bibinfo {title} {Convective
  mixing by internal waves in the {P}uerto {R}ico {T}rench},}\ }\href@noop {}
  {\bibfield  {journal} {\bibinfo  {journal} {J. Mar. Res.}\ }\textbf {\bibinfo
  {volume} {74}},\ \bibinfo {pages} {161--173} (\bibinfo {year}
  {2016})}\BibitemShut {NoStop}%
\bibitem [{\citenamefont {Capet}\ \emph {et~al.}(2008)\citenamefont {Capet},
  \citenamefont {McWilliams}, \citenamefont {Molemaker},\ and\ \citenamefont
  {Shchepetkin}}]{capet}%
  \BibitemOpen
  \bibfield  {author} {\bibinfo {author} {\bibfnamefont {X.}~\bibnamefont
  {Capet}}, \bibinfo {author} {\bibfnamefont {J.}~\bibnamefont {McWilliams}},
  \bibinfo {author} {\bibfnamefont {M.}~\bibnamefont {Molemaker}}, \ and\
  \bibinfo {author} {\bibfnamefont {A.}~\bibnamefont {Shchepetkin}},\
  }\bibfield  {title} {\enquote {\bibinfo {title} {Mesoscale to submesoscale
  transition in the california current system. part i: Flow structure, eddy
  flux, and observational tests},}\ }\href@noop {} {\bibfield  {journal}
  {\bibinfo  {journal} {J. Phys. Oceanogr.}\ }\textbf {\bibinfo {volume}
  {38}},\ \bibinfo {pages} {29--43} (\bibinfo {year} {2008})}\BibitemShut
  {NoStop}%
\bibitem [{\citenamefont {Klymak}\ \emph {et~al.}(2015)\citenamefont {Klymak},
  \citenamefont {Crawford}, \citenamefont {Alford}, \citenamefont {MacKinnon},\
  and\ \citenamefont {Pinkel}}]{klymak_15}%
  \BibitemOpen
  \bibfield  {author} {\bibinfo {author} {\bibfnamefont {J.M.}\ \bibnamefont
  {Klymak}}, \bibinfo {author} {\bibfnamefont {W.}~\bibnamefont {Crawford}},
  \bibinfo {author} {\bibfnamefont {M.H.}\ \bibnamefont {Alford}}, \bibinfo
  {author} {\bibfnamefont {J.A.}\ \bibnamefont {MacKinnon}}, \ and\ \bibinfo
  {author} {\bibfnamefont {R.}~\bibnamefont {Pinkel}},\ }\bibfield  {title}
  {\enquote {\bibinfo {title} {Along-isopycnal variability of spice in the
  {N}orth {P}acific},}\ }\href@noop {} {\bibfield  {journal} {\bibinfo
  {journal} {J. Geophys. Res.}\ }\textbf {\bibinfo {volume} {120}},\ \bibinfo
  {pages} {2287--2307} (\bibinfo {year} {2015})}\BibitemShut {NoStop}%
\bibitem [{\citenamefont {Riley}\ \emph {et~al.}(1981)\citenamefont {Riley},
  \citenamefont {Metcalfe},\ and\ \citenamefont {Weissman}}]{Riley_1981}%
  \BibitemOpen
  \bibfield  {author} {\bibinfo {author} {\bibfnamefont {J.J.}\ \bibnamefont
  {Riley}}, \bibinfo {author} {\bibfnamefont {R.W.}\ \bibnamefont {Metcalfe}},
  \ and\ \bibinfo {author} {\bibfnamefont {M.A.}\ \bibnamefont {Weissman}},\
  }\bibfield  {title} {\enquote {\bibinfo {title} {Direct numerical simulations
  of homogeneous turbulence in density-stratified fluids},}\ }\href@noop {}
  {\bibfield  {journal} {\bibinfo  {journal} {AIP Conference Proceedings}\
  }\textbf {\bibinfo {volume} {76}},\ \bibinfo {pages} {79--112} (\bibinfo
  {year} {1981})}\BibitemShut {NoStop}%
\bibitem [{\citenamefont {Herring}\ and\ \citenamefont
  {M{\'e}tais}(1989)}]{Herring_1989}%
  \BibitemOpen
  \bibfield  {author} {\bibinfo {author} {\bibfnamefont {J.R.}\ \bibnamefont
  {Herring}}\ and\ \bibinfo {author} {\bibfnamefont {O.}~\bibnamefont
  {M{\'e}tais}},\ }\bibfield  {title} {\enquote {\bibinfo {title} {Numerical
  experiments in forced stably stratified turbulence},}\ }\href@noop {}
  {\bibfield  {journal} {\bibinfo  {journal} {J. Fluid Mech.}\ }\textbf
  {\bibinfo {volume} {202}},\ \bibinfo {pages} {97--115} (\bibinfo {year}
  {1989})}\BibitemShut {NoStop}%
\bibitem [{\citenamefont {Winters}\ \emph {et~al.}(1995)\citenamefont
  {Winters}, \citenamefont {Lombard}, \citenamefont {Riley},\ and\
  \citenamefont {D'Asaro}}]{Winters_1995}%
  \BibitemOpen
  \bibfield  {author} {\bibinfo {author} {\bibfnamefont {K.B.}\ \bibnamefont
  {Winters}}, \bibinfo {author} {\bibfnamefont {P.N.}\ \bibnamefont {Lombard}},
  \bibinfo {author} {\bibfnamefont {J.J.}\ \bibnamefont {Riley}}, \ and\
  \bibinfo {author} {\bibfnamefont {E.}~\bibnamefont {D'Asaro}},\ }\bibfield
  {title} {\enquote {\bibinfo {title} {Available potential energy and mixing in
  density-stratified fluids},}\ }\href@noop {} {\bibfield  {journal} {\bibinfo
  {journal} {J. Fluid Mech.}\ }\textbf {\bibinfo {volume} {289}},\ \bibinfo
  {pages} {115--128} (\bibinfo {year} {1995})}\BibitemShut {NoStop}%
\bibitem [{\citenamefont {M{\'e}tais}\ \emph {et~al.}(1996)\citenamefont
  {M{\'e}tais}, \citenamefont {Bartello}, \citenamefont {Garnier},
  \citenamefont {Riley},\ and\ \citenamefont {Lesieur}}]{Metais_1996}%
  \BibitemOpen
  \bibfield  {author} {\bibinfo {author} {\bibfnamefont {O.}~\bibnamefont
  {M{\'e}tais}}, \bibinfo {author} {\bibfnamefont {P.}~\bibnamefont
  {Bartello}}, \bibinfo {author} {\bibfnamefont {E.}~\bibnamefont {Garnier}},
  \bibinfo {author} {\bibfnamefont {J.J.}\ \bibnamefont {Riley}}, \ and\
  \bibinfo {author} {\bibfnamefont {M.}~\bibnamefont {Lesieur}},\ }\bibfield
  {title} {\enquote {\bibinfo {title} {Inverse cascade in stably stratified
  rotating turbulence},}\ }\href@noop {} {\bibfield  {journal} {\bibinfo
  {journal} {Dynamics of Atmospheres and Oceans}\ }\textbf {\bibinfo {volume}
  {23}},\ \bibinfo {pages} {193--203} (\bibinfo {year} {1996})}\BibitemShut
  {NoStop}%
\bibitem [{\citenamefont {Kimura}\ and\ \citenamefont
  {Herring}(1996)}]{Kimura_1996}%
  \BibitemOpen
  \bibfield  {author} {\bibinfo {author} {\bibfnamefont {Y.}~\bibnamefont
  {Kimura}}\ and\ \bibinfo {author} {\bibfnamefont {J.R.}\ \bibnamefont
  {Herring}},\ }\bibfield  {title} {\enquote {\bibinfo {title} {Diffusion in
  stably stratified turbulence},}\ }\href@noop {} {\bibfield  {journal}
  {\bibinfo  {journal} {J. Fluid Mech.}\ }\textbf {\bibinfo {volume} {328}},\
  \bibinfo {pages} {253--269} (\bibinfo {year} {1996})}\BibitemShut {NoStop}%
\bibitem [{\citenamefont {Billant}\ and\ \citenamefont
  {Chomaz}(2000)}]{Billant_2000}%
  \BibitemOpen
  \bibfield  {author} {\bibinfo {author} {\bibfnamefont {P.}~\bibnamefont
  {Billant}}\ and\ \bibinfo {author} {\bibfnamefont {J-M.}\ \bibnamefont
  {Chomaz}},\ }\bibfield  {title} {\enquote {\bibinfo {title} {Experimental
  evidence for a new instability of a vertical columnar vortex pair in a
  strongly stratified fluid},}\ }\href@noop {} {\bibfield  {journal} {\bibinfo
  {journal} {J. Fluid Mech.}\ }\textbf {\bibinfo {volume} {418}},\ \bibinfo
  {pages} {167--188} (\bibinfo {year} {2000})}\BibitemShut {NoStop}%
\bibitem [{\citenamefont {Riley}\ and\ \citenamefont
  {de~Bruyn~Kops}(2003)}]{Riley_2003}%
  \BibitemOpen
  \bibfield  {author} {\bibinfo {author} {\bibfnamefont {J.J.}\ \bibnamefont
  {Riley}}\ and\ \bibinfo {author} {\bibfnamefont {S.M.}\ \bibnamefont
  {de~Bruyn~Kops}},\ }\bibfield  {title} {\enquote {\bibinfo {title} {Dynamics
  of turbulence strongly influenced by buoyancy},}\ }\href@noop {} {\bibfield
  {journal} {\bibinfo  {journal} {Physics of Fluids}\ }\textbf {\bibinfo
  {volume} {15}},\ \bibinfo {pages} {2047--2059} (\bibinfo {year}
  {2003})}\BibitemShut {NoStop}%
\bibitem [{\citenamefont {Lindborg}(2006)}]{Lindborg_2006}%
  \BibitemOpen
  \bibfield  {author} {\bibinfo {author} {\bibfnamefont {E.}~\bibnamefont
  {Lindborg}},\ }\bibfield  {title} {\enquote {\bibinfo {title} {The energy
  cascade in a strongly stratified fluid},}\ }\href@noop {} {\bibfield
  {journal} {\bibinfo  {journal} {J. Fluid Mech.}\ }\textbf {\bibinfo {volume}
  {550}},\ \bibinfo {pages} {207--242} (\bibinfo {year} {2006})}\BibitemShut
  {NoStop}%
\bibitem [{\citenamefont {Laval}\ \emph {et~al.}(2003)\citenamefont {Laval},
  \citenamefont {McWilliams},\ and\ \citenamefont {Dubrulle}}]{Laval_2003}%
  \BibitemOpen
  \bibfield  {author} {\bibinfo {author} {\bibfnamefont {J-P.}\ \bibnamefont
  {Laval}}, \bibinfo {author} {\bibfnamefont {J.C.}\ \bibnamefont
  {McWilliams}}, \ and\ \bibinfo {author} {\bibfnamefont {B.}~\bibnamefont
  {Dubrulle}},\ }\bibfield  {title} {\enquote {\bibinfo {title} {Forced
  stratified turbulence: successive transitions with {R}eynolds number},}\
  }\href@noop {} {\bibfield  {journal} {\bibinfo  {journal} {Physical Review
  E}\ }\textbf {\bibinfo {volume} {68}},\ \bibinfo {pages} {036308} (\bibinfo
  {year} {2003})}\BibitemShut {NoStop}%
\bibitem [{\citenamefont {Bartello}\ and\ \citenamefont
  {Tobias}(2013)}]{Bartello_2013}%
  \BibitemOpen
  \bibfield  {author} {\bibinfo {author} {\bibfnamefont {P.}~\bibnamefont
  {Bartello}}\ and\ \bibinfo {author} {\bibfnamefont {S.M.}\ \bibnamefont
  {Tobias}},\ }\bibfield  {title} {\enquote {\bibinfo {title} {Sensitivity of
  stratified turbulence to the buoyancy {R}eynolds number},}\ }\href@noop {}
  {\bibfield  {journal} {\bibinfo  {journal} {J. Fluid Mech.}\ }\textbf
  {\bibinfo {volume} {725}},\ \bibinfo {pages} {1--22} (\bibinfo {year}
  {2013})}\BibitemShut {NoStop}%
\bibitem [{\citenamefont {Ivey}\ \emph {et~al.}(2008)\citenamefont {Ivey},
  \citenamefont {Winters},\ and\ \citenamefont {Koseff}}]{ivey_08}%
  \BibitemOpen
  \bibfield  {author} {\bibinfo {author} {\bibfnamefont {G.}~\bibnamefont
  {Ivey}}, \bibinfo {author} {\bibfnamefont {K.}~\bibnamefont {Winters}}, \
  and\ \bibinfo {author} {\bibfnamefont {J.}~\bibnamefont {Koseff}},\
  }\bibfield  {title} {\enquote {\bibinfo {title} {Density stratification,
  turbulence but how much mixing?}}\ }\href@noop {} {\bibfield  {journal}
  {\bibinfo  {journal} {Ann. Rev. Fluid Mech.}\ }\textbf {\bibinfo {volume}
  {{\bf 40}}},\ \bibinfo {pages} {169--184} (\bibinfo {year}
  {2008})}\BibitemShut {NoStop}%
\bibitem [{\citenamefont {Brethouwer}\ \emph {et~al.}(2007)\citenamefont
  {Brethouwer}, \citenamefont {Billant}, \citenamefont {Lindborg},\ and\
  \citenamefont {Chomaz}}]{Brethouwer_2007}%
  \BibitemOpen
  \bibfield  {author} {\bibinfo {author} {\bibfnamefont {G.}~\bibnamefont
  {Brethouwer}}, \bibinfo {author} {\bibfnamefont {P.}~\bibnamefont {Billant}},
  \bibinfo {author} {\bibfnamefont {E.}~\bibnamefont {Lindborg}}, \ and\
  \bibinfo {author} {\bibfnamefont {J-M.}\ \bibnamefont {Chomaz}},\ }\bibfield
  {title} {\enquote {\bibinfo {title} {Scaling analysis and simulation of
  strongly stratified turbulent flows},}\ }\href@noop {} {\bibfield  {journal}
  {\bibinfo  {journal} {J. Fluid Mech.}\ }\textbf {\bibinfo {volume} {585}},\
  \bibinfo {pages} {343--368} (\bibinfo {year} {2007})}\BibitemShut {NoStop}%
\bibitem [{\citenamefont {Davidson}(2013)}]{davidson}%
  \BibitemOpen
  \bibfield  {author} {\bibinfo {author} {\bibfnamefont {P.A.}\ \bibnamefont
  {Davidson}},\ }\href@noop {} {\emph {\bibinfo {title} {Turbulence in
  rotating, stratified and electrically conducting fluids}}}\ (\bibinfo
  {publisher} {Cambridge University Press},\ \bibinfo {year}
  {2013})\BibitemShut {NoStop}%
\bibitem [{\citenamefont {Pouquet}\ \emph
  {et~al.}(2019{\natexlab{c}})\citenamefont {Pouquet}, \citenamefont
  {Rosenberg},\ and\ \citenamefont {Marino}}]{pouquet_19p}%
  \BibitemOpen
  \bibfield  {author} {\bibinfo {author} {\bibfnamefont {A.}~\bibnamefont
  {Pouquet}}, \bibinfo {author} {\bibfnamefont {D.}~\bibnamefont {Rosenberg}},
  \ and\ \bibinfo {author} {\bibfnamefont {R.}~\bibnamefont {Marino}},\
  }\bibfield  {title} {\enquote {\bibinfo {title} {Linking dissipation,
  anisotropy and intermittency in rotating stratified turbulence},}\
  }\href@noop {} {\bibfield  {journal} {\bibinfo  {journal} {Phys. Fluids}\
  }\textbf {\bibinfo {volume} {31}},\ \bibinfo {pages} {105116} (\bibinfo
  {year} {2019}{\natexlab{c}})}\BibitemShut {NoStop}%
\bibitem [{\citenamefont {Rorai}\ \emph {et~al.}(2014)\citenamefont {Rorai},
  \citenamefont {Mininni},\ and\ \citenamefont {Pouquet}}]{rorai2014}%
  \BibitemOpen
  \bibfield  {author} {\bibinfo {author} {\bibfnamefont {C.}~\bibnamefont
  {Rorai}}, \bibinfo {author} {\bibfnamefont {P.D.}\ \bibnamefont {Mininni}}, \
  and\ \bibinfo {author} {\bibfnamefont {A.}~\bibnamefont {Pouquet}},\
  }\bibfield  {title} {\enquote {\bibinfo {title} {Turbulence comes in bursts
  in stably stratified flows},}\ }\href@noop {} {\bibfield  {journal} {\bibinfo
   {journal} {Phys. Rev. E}\ }\textbf {\bibinfo {volume} {89}},\ \bibinfo
  {pages} {043002} (\bibinfo {year} {2014})}\BibitemShut {NoStop}%
\bibitem [{\citenamefont {de~Bruyn~Kops}(2015)}]{deBruyn_2015}%
  \BibitemOpen
  \bibfield  {author} {\bibinfo {author} {\bibfnamefont {S.M.}\ \bibnamefont
  {de~Bruyn~Kops}},\ }\bibfield  {title} {\enquote {\bibinfo {title} {Classical
  scaling and intermittency in strongly stratified {B}oussinesq turbulence},}\
  }\href@noop {} {\bibfield  {journal} {\bibinfo  {journal} {J. Fluid Mech.}\
  }\textbf {\bibinfo {volume} {775}},\ \bibinfo {pages} {436--463} (\bibinfo
  {year} {2015})}\BibitemShut {NoStop}%
\bibitem [{\citenamefont {Feraco}\ \emph {et~al.}(2018)\citenamefont {Feraco},
  \citenamefont {Marino}, \citenamefont {Pumir}, \citenamefont {Primavera},
  \citenamefont {Mininni}, \citenamefont {Pouquet},\ and\ \citenamefont
  {Rosenberg}}]{feraco2018}%
  \BibitemOpen
  \bibfield  {author} {\bibinfo {author} {\bibfnamefont {F.}~\bibnamefont
  {Feraco}}, \bibinfo {author} {\bibfnamefont {R.}~\bibnamefont {Marino}},
  \bibinfo {author} {\bibfnamefont {A.}~\bibnamefont {Pumir}}, \bibinfo
  {author} {\bibfnamefont {L.}~\bibnamefont {Primavera}}, \bibinfo {author}
  {\bibfnamefont {P.D.}\ \bibnamefont {Mininni}}, \bibinfo {author}
  {\bibfnamefont {A.}~\bibnamefont {Pouquet}}, \ and\ \bibinfo {author}
  {\bibfnamefont {D.}~\bibnamefont {Rosenberg}},\ }\bibfield  {title} {\enquote
  {\bibinfo {title} {Vertical drafts and mixing in stratified turbulence: sharp
  transition with {F}roude number},}\ }\href@noop {} {\bibfield  {journal}
  {\bibinfo  {journal} {Eur. Phys. Lett.}\ }\textbf {\bibinfo {volume} {123}},\
  \bibinfo {pages} {44002} (\bibinfo {year} {2018})}\BibitemShut {NoStop}%
\bibitem [{\citenamefont {Smyth}\ \emph {et~al.}(2019)\citenamefont {Smyth},
  \citenamefont {Nash},\ and\ \citenamefont {Moum}}]{Smyth_2019}%
  \BibitemOpen
  \bibfield  {author} {\bibinfo {author} {\bibfnamefont {W.D.}\ \bibnamefont
  {Smyth}}, \bibinfo {author} {\bibfnamefont {J.D.}\ \bibnamefont {Nash}}, \
  and\ \bibinfo {author} {\bibfnamefont {J.N.}\ \bibnamefont {Moum}},\
  }\bibfield  {title} {\enquote {\bibinfo {title} {Self-organized criticality
  in geophysical turbulence},}\ }\href@noop {} {\bibfield  {journal} {\bibinfo
  {journal} {Scientific reports}\ }\textbf {\bibinfo {volume} {9}},\ \bibinfo
  {pages} {3747} (\bibinfo {year} {2019})}\BibitemShut {NoStop}%
\bibitem [{\citenamefont {Feraco}\ \emph {et~al.}(2021)\citenamefont {Feraco},
  \citenamefont {Marino}, \citenamefont {Primavera}, \citenamefont {Pumir},
  \citenamefont {Mininni}, \citenamefont {Rosenberg}, \citenamefont {Pouquet},
  \citenamefont {Foldes}, \citenamefont {L{\'e}v{\^e}que}, \citenamefont
  {Camporeale}, \citenamefont {Cerri}, \citenamefont {Asokan}, \citenamefont
  {Chau}, \citenamefont {Bertoglio}, \citenamefont {Salizzoni},\ and\
  \citenamefont {Marro}}]{feraco2021}%
  \BibitemOpen
  \bibfield  {author} {\bibinfo {author} {\bibfnamefont {F.}~\bibnamefont
  {Feraco}}, \bibinfo {author} {\bibfnamefont {R.}~\bibnamefont {Marino}},
  \bibinfo {author} {\bibfnamefont {L.}~\bibnamefont {Primavera}}, \bibinfo
  {author} {\bibfnamefont {A.}~\bibnamefont {Pumir}}, \bibinfo {author}
  {\bibfnamefont {P.D.}\ \bibnamefont {Mininni}}, \bibinfo {author}
  {\bibfnamefont {D.}~\bibnamefont {Rosenberg}}, \bibinfo {author}
  {\bibfnamefont {A.}~\bibnamefont {Pouquet}}, \bibinfo {author} {\bibfnamefont
  {R.}~\bibnamefont {Foldes}}, \bibinfo {author} {\bibfnamefont
  {E.}~\bibnamefont {L{\'e}v{\^e}que}}, \bibinfo {author} {\bibfnamefont
  {E.}~\bibnamefont {Camporeale}}, \bibinfo {author} {\bibfnamefont {S.S.}\
  \bibnamefont {Cerri}}, \bibinfo {author} {\bibfnamefont {H.~Charuvil}\
  \bibnamefont {Asokan}}, \bibinfo {author} {\bibfnamefont {J.L.}\ \bibnamefont
  {Chau}}, \bibinfo {author} {\bibfnamefont {J.P.}\ \bibnamefont {Bertoglio}},
  \bibinfo {author} {\bibfnamefont {P.}~\bibnamefont {Salizzoni}}, \ and\
  \bibinfo {author} {\bibfnamefont {M.}~\bibnamefont {Marro}},\ }\bibfield
  {title} {\enquote {\bibinfo {title} {Connecting large-scale velocity and
  temperature bursts with small-scale intermittency in stratified
  turbulence},}\ }\href@noop {} {\bibfield  {journal} {\bibinfo  {journal}
  {Eur. Phys. Lett.}\ }\textbf {\bibinfo {volume} {135}},\ \bibinfo {pages}
  {14001} (\bibinfo {year} {2021})}\BibitemShut {NoStop}%
\bibitem [{\citenamefont {Petoukhov}\ \emph {et~al.}(2008)\citenamefont
  {Petoukhov}, \citenamefont {Eliseev}, \citenamefont {Klein},\ and\
  \citenamefont {Oesterle}}]{petoukhov_08}%
  \BibitemOpen
  \bibfield  {author} {\bibinfo {author} {\bibfnamefont {V.}~\bibnamefont
  {Petoukhov}}, \bibinfo {author} {\bibfnamefont {A.V.}\ \bibnamefont
  {Eliseev}}, \bibinfo {author} {\bibfnamefont {R.}~\bibnamefont {Klein}}, \
  and\ \bibinfo {author} {\bibfnamefont {H.}~\bibnamefont {Oesterle}},\
  }\bibfield  {title} {\enquote {\bibinfo {title} {On statistics of the
  free-troposphere synoptic component: an evaluation of skewnesses and mixed
  third-order moments contribution to the synoptic-scale dynamics and fluxes of
  heat and humidity},}\ }\href@noop {} {\bibfield  {journal} {\bibinfo
  {journal} {Tellus}\ }\textbf {\bibinfo {volume} {60A}},\ \bibinfo {pages}
  {11--31} (\bibinfo {year} {2008})}\BibitemShut {NoStop}%
\bibitem [{\citenamefont {Sardeshmukh}\ \emph {et~al.}(2015)\citenamefont
  {Sardeshmukh}, \citenamefont {Compo},\ and\ \citenamefont
  {Penland}}]{sardeshmukh_15}%
  \BibitemOpen
  \bibfield  {author} {\bibinfo {author} {\bibfnamefont {P.}~\bibnamefont
  {Sardeshmukh}}, \bibinfo {author} {\bibfnamefont {G.P.}\ \bibnamefont
  {Compo}}, \ and\ \bibinfo {author} {\bibfnamefont {C.}~\bibnamefont
  {Penland}},\ }\bibfield  {title} {\enquote {\bibinfo {title} {Need for
  caution in interpreting extreme weather statistics},}\ }\href@noop {}
  {\bibfield  {journal} {\bibinfo  {journal} {J. Climate}\ }\textbf {\bibinfo
  {volume} {28}},\ \bibinfo {pages} {9166--9185} (\bibinfo {year}
  {2015})}\BibitemShut {NoStop}%
\bibitem [{\citenamefont {Sujovolsky}\ and\ \citenamefont
  {Mininni}(2020)}]{sujovolsky_20}%
  \BibitemOpen
  \bibfield  {author} {\bibinfo {author} {\bibfnamefont {N.E.}\ \bibnamefont
  {Sujovolsky}}\ and\ \bibinfo {author} {\bibfnamefont {P.D.}\ \bibnamefont
  {Mininni}},\ }\bibfield  {title} {\enquote {\bibinfo {title} {From waves to
  convection and back again: The phase space of stably stratified
  turbulence},}\ }\href@noop {} {\bibfield  {journal} {\bibinfo  {journal}
  {Phys. Rev. F}\ }\textbf {\bibinfo {volume} {5}},\ \bibinfo {pages} {064802}
  (\bibinfo {year} {2020})}\BibitemShut {NoStop}%
\bibitem [{\citenamefont {Salehipour}\ \emph {et~al.}(2016)\citenamefont
  {Salehipour}, \citenamefont {Peltier}, \citenamefont {Whalen},\ and\
  \citenamefont {MacKinnon}}]{salehipour_16}%
  \BibitemOpen
  \bibfield  {author} {\bibinfo {author} {\bibfnamefont {H.}~\bibnamefont
  {Salehipour}}, \bibinfo {author} {\bibfnamefont {W.R.}\ \bibnamefont
  {Peltier}}, \bibinfo {author} {\bibfnamefont {C.B.}\ \bibnamefont {Whalen}},
  \ and\ \bibinfo {author} {\bibfnamefont {J.A.}\ \bibnamefont {MacKinnon}},\
  }\bibfield  {title} {\enquote {\bibinfo {title} {A new characterization of
  the turbulent diapycnal diffusivities of mass and momentum in the ocean},}\
  }\href@noop {} {\bibfield  {journal} {\bibinfo  {journal} {Geophys. Res.
  Lett.}\ }\textbf {\bibinfo {volume} {43}},\ \bibinfo {pages} {3370--3379}
  (\bibinfo {year} {2016})}\BibitemShut {NoStop}%
\bibitem [{\citenamefont {Vallis}(2017)}]{Vallis}%
  \BibitemOpen
  \bibfield  {author} {\bibinfo {author} {\bibfnamefont {G.K.}\ \bibnamefont
  {Vallis}},\ }\href@noop {} {\emph {\bibinfo {title} {Atmospheric and oceanic
  fluid dynamics}}}\ (\bibinfo  {publisher} {Cambridge University Press},\
  \bibinfo {year} {2017})\BibitemShut {NoStop}%
\bibitem [{\citenamefont {Mininni}\ \emph {et~al.}(2011)\citenamefont
  {Mininni}, \citenamefont {Rosenberg}, \citenamefont {Reddy},\ and\
  \citenamefont {Pouquet}}]{mininni11}%
  \BibitemOpen
  \bibfield  {author} {\bibinfo {author} {\bibfnamefont {P.D.}\ \bibnamefont
  {Mininni}}, \bibinfo {author} {\bibfnamefont {D.}~\bibnamefont {Rosenberg}},
  \bibinfo {author} {\bibfnamefont {R.}~\bibnamefont {Reddy}}, \ and\ \bibinfo
  {author} {\bibfnamefont {A.}~\bibnamefont {Pouquet}},\ }\bibfield  {title}
  {\enquote {\bibinfo {title} {A hybrid {M}{P}{I}-{O}pen{M}{P} scheme for
  scalable parallel pseudospectral computations for fluid turbulence},}\
  }\href@noop {} {\bibfield  {journal} {\bibinfo  {journal} {Parallel
  Computing}\ }\textbf {\bibinfo {volume} {37}},\ \bibinfo {pages} {316--326}
  (\bibinfo {year} {2011})}\BibitemShut {NoStop}%
\bibitem [{\citenamefont {Fontana}\ \emph {et~al.}(2020)\citenamefont
  {Fontana}, \citenamefont {Bruno}, \citenamefont {Mininni},\ and\
  \citenamefont {Dmitruk}}]{fontana_20}%
  \BibitemOpen
  \bibfield  {author} {\bibinfo {author} {\bibfnamefont {M.}~\bibnamefont
  {Fontana}}, \bibinfo {author} {\bibfnamefont {O.P.}\ \bibnamefont {Bruno}},
  \bibinfo {author} {\bibfnamefont {P.D.}\ \bibnamefont {Mininni}}, \ and\
  \bibinfo {author} {\bibfnamefont {P.}~\bibnamefont {Dmitruk}},\ }\bibfield
  {title} {\enquote {\bibinfo {title} {Fourier continuation method for
  incompressible fluids with boundaries},}\ }\href@noop {} {\bibfield
  {journal} {\bibinfo  {journal} {Comp. Phys. Comm.}\ }\textbf {\bibinfo
  {volume} {256}},\ \bibinfo {pages} {107482} (\bibinfo {year}
  {2020})}\BibitemShut {NoStop}%
\bibitem [{\citenamefont {Rosenberg}\ \emph {et~al.}(2020)\citenamefont
  {Rosenberg}, \citenamefont {Mininni}, \citenamefont {Reddy},\ and\
  \citenamefont {Pouquet}}]{rosenberg_20}%
  \BibitemOpen
  \bibfield  {author} {\bibinfo {author} {\bibfnamefont {D.}~\bibnamefont
  {Rosenberg}}, \bibinfo {author} {\bibfnamefont {P.D.}\ \bibnamefont
  {Mininni}}, \bibinfo {author} {\bibfnamefont {R.}~\bibnamefont {Reddy}}, \
  and\ \bibinfo {author} {\bibfnamefont {A.}~\bibnamefont {Pouquet}},\
  }\bibfield  {title} {\enquote {\bibinfo {title} {{G}{P}{U} parallelization of
  a hybrid pseudospectral geophysical turbulence framework using
  {C}{U}{D}{A}},}\ }\href@noop {} {\bibfield  {journal} {\bibinfo  {journal}
  {Atmosphere}\ }\textbf {\bibinfo {volume} {11}},\ \bibinfo {pages} {178}
  (\bibinfo {year} {2020})}\BibitemShut {NoStop}%
\bibitem [{\citenamefont {Pumir}(1996)}]{pumir96}%
  \BibitemOpen
  \bibfield  {author} {\bibinfo {author} {\bibfnamefont {A.}~\bibnamefont
  {Pumir}},\ }\bibfield  {title} {\enquote {\bibinfo {title} {Turbulence in
  homogeneous shear flows},}\ }\href@noop {} {\bibfield  {journal} {\bibinfo
  {journal} {Phys. Fluids}\ }\textbf {\bibinfo {volume} {8}},\ \bibinfo {pages}
  {3112--3127} (\bibinfo {year} {1996})}\BibitemShut {NoStop}%
\bibitem [{\citenamefont {Sekimoto}\ \emph {et~al.}(2016)\citenamefont
  {Sekimoto}, \citenamefont {Dong},\ and\ \citenamefont
  {Jim{\'e}nez}}]{sekimoto16}%
  \BibitemOpen
  \bibfield  {author} {\bibinfo {author} {\bibfnamefont {A.}~\bibnamefont
  {Sekimoto}}, \bibinfo {author} {\bibfnamefont {S.}~\bibnamefont {Dong}}, \
  and\ \bibinfo {author} {\bibfnamefont {J.}~\bibnamefont {Jim{\'e}nez}},\
  }\bibfield  {title} {\enquote {\bibinfo {title} {Direct numerical simulation
  of statistically stationary and homogeneous shear turbulence and its relation
  to other shear flows},}\ }\href@noop {} {\bibfield  {journal} {\bibinfo
  {journal} {Phys. Fluids}\ }\textbf {\bibinfo {volume} {28}},\ \bibinfo
  {pages} {035101} (\bibinfo {year} {2016})}\BibitemShut {NoStop}%
\bibitem [{\citenamefont {Sz{\'e}kely}\ \emph {et~al.}(2007)\citenamefont
  {Sz{\'e}kely}, \citenamefont {Rizzo},\ and\ \citenamefont
  {Bakirov}}]{szekely}%
  \BibitemOpen
  \bibfield  {author} {\bibinfo {author} {\bibfnamefont {G.J.}\ \bibnamefont
  {Sz{\'e}kely}}, \bibinfo {author} {\bibfnamefont {M.L.}\ \bibnamefont
  {Rizzo}}, \ and\ \bibinfo {author} {\bibfnamefont {N.K.}\ \bibnamefont
  {Bakirov}},\ }\bibfield  {title} {\enquote {\bibinfo {title} {Measuring and
  testing dependence by correlation of distances},}\ }\href@noop {} {\bibfield
  {journal} {\bibinfo  {journal} {The Annals of Statistics}\ }\textbf {\bibinfo
  {volume} {35}},\ \bibinfo {pages} {2769--2794} (\bibinfo {year}
  {2007})}\BibitemShut {NoStop}%
\bibitem [{\citenamefont {Edelmann}\ \emph {et~al.}(2021)\citenamefont
  {Edelmann}, \citenamefont {M{\'o}ri},\ and\ \citenamefont
  {Sz{\'e}kely}}]{edelmann}%
  \BibitemOpen
  \bibfield  {author} {\bibinfo {author} {\bibfnamefont {D.}~\bibnamefont
  {Edelmann}}, \bibinfo {author} {\bibfnamefont {T.F.}\ \bibnamefont
  {M{\'o}ri}}, \ and\ \bibinfo {author} {\bibfnamefont {G.J.}\ \bibnamefont
  {Sz{\'e}kely}},\ }\bibfield  {title} {\enquote {\bibinfo {title} {On
  relationships between the {P}earson and the distance correlation
  coefficients},}\ }\href@noop {} {\bibfield  {journal} {\bibinfo  {journal}
  {Statistics \& Probability Letters}\ }\textbf {\bibinfo {volume} {169}},\
  \bibinfo {pages} {108960} (\bibinfo {year} {2021})}\BibitemShut {NoStop}%
\bibitem [{SM()}]{SM}%
  \BibitemOpen
  \href@noop {} {\bibinfo  {journal} {See Supplemental Material at [URL will be
  inserted by publisher] for a movie showing the time evolution of typical
  extreme events developing in run P5. In the left panel of the movie it is
  rendered one of the horizontal components of the velocity ($u$), while the
  right panel displays the values of the vertical velocity ($w$) exceeding four
  standard deviations, namely the extreme vertical drafts corresponding to
  $w>4\sigma_w$. The movie shows the time evolution of the standardized flow
  fields over an interval of $\sim 26\tau_{NL}$}\ }\BibitemShut {NoStop}%
\bibitem [{\citenamefont {Pumir}(1994)}]{pumir_94}%
  \BibitemOpen
\bibfield  {journal} {  }\bibfield  {author} {\bibinfo {author} {\bibfnamefont
  {A.}~\bibnamefont {Pumir}},\ }\bibfield  {title} {\enquote {\bibinfo {title}
  {A numerical study of the mixing of a passive scalar in three dimensions in
  the presence of a mean gradient},}\ }\href@noop {} {\bibfield  {journal}
  {\bibinfo  {journal} {Phys. Fluids}\ }\textbf {\bibinfo {volume} {6}},\
  \bibinfo {pages} {2118--2132} (\bibinfo {year} {1994})}\BibitemShut {NoStop}%
\bibitem [{\citenamefont {Sujovolsky}\ \emph {et~al.}(2018)\citenamefont
  {Sujovolsky}, \citenamefont {Mininni},\ and\ \citenamefont
  {Pouquet}}]{sujovolsky_18}%
  \BibitemOpen
  \bibfield  {author} {\bibinfo {author} {\bibfnamefont {N.E.}\ \bibnamefont
  {Sujovolsky}}, \bibinfo {author} {\bibfnamefont {P.D.}\ \bibnamefont
  {Mininni}}, \ and\ \bibinfo {author} {\bibfnamefont {A.}~\bibnamefont
  {Pouquet}},\ }\bibfield  {title} {\enquote {\bibinfo {title} {Generation of
  turbulence through frontogenesis in sheared stratified flows},}\ }\href@noop
  {} {\bibfield  {journal} {\bibinfo  {journal} {Phys. Fluids}\ }\textbf
  {\bibinfo {volume} {30}},\ \bibinfo {pages} {086601} (\bibinfo {year}
  {2018})}\BibitemShut {NoStop}%
\bibitem [{\citenamefont {Vieillefosse}(1984)}]{Vieillefosse_1984}%
  \BibitemOpen
  \bibfield  {author} {\bibinfo {author} {\bibfnamefont {P}~\bibnamefont
  {Vieillefosse}},\ }\bibfield  {title} {\enquote {\bibinfo {title} {Internal
  motion of a small element of fluid in an inviscid flow},}\ }\href@noop {}
  {\bibfield  {journal} {\bibinfo  {journal} {Physica A: Statistical Mechanics
  and its Applications}\ }\textbf {\bibinfo {volume} {125}},\ \bibinfo {pages}
  {150--162} (\bibinfo {year} {1984})}\BibitemShut {NoStop}%
\bibitem [{\citenamefont {Meneveau}(2011)}]{Meneveau_2011}%
  \BibitemOpen
  \bibfield  {author} {\bibinfo {author} {\bibfnamefont {Charles}\ \bibnamefont
  {Meneveau}},\ }\bibfield  {title} {\enquote {\bibinfo {title} {Lagrangian
  dynamics and models of the velocity gradient tensor in turbulent flows},}\
  }\href@noop {} {\bibfield  {journal} {\bibinfo  {journal} {Annual Review of
  Fluid Mechanics}\ }\textbf {\bibinfo {volume} {43}},\ \bibinfo {pages}
  {219--245} (\bibinfo {year} {2011})}\BibitemShut {NoStop}%
\bibitem [{\citenamefont {Sujovolsky}\ \emph {et~al.}(2019)\citenamefont
  {Sujovolsky}, \citenamefont {Mindlin},\ and\ \citenamefont
  {Mininni}}]{Sujovolsky_2019}%
  \BibitemOpen
  \bibfield  {author} {\bibinfo {author} {\bibfnamefont {N.~E.}\ \bibnamefont
  {Sujovolsky}}, \bibinfo {author} {\bibfnamefont {G.~B.}\ \bibnamefont
  {Mindlin}}, \ and\ \bibinfo {author} {\bibfnamefont {P.~D.}\ \bibnamefont
  {Mininni}},\ }\bibfield  {title} {\enquote {\bibinfo {title} {Invariant
  manifolds in stratified turbulence},}\ }\href@noop {} {\bibfield  {journal}
  {\bibinfo  {journal} {Physical Review Fluids}\ }\textbf {\bibinfo {volume}
  {4}},\ \bibinfo {pages} {052402(R)} (\bibinfo {year} {2019})}\BibitemShut
  {NoStop}%
\bibitem [{\citenamefont {Pomeau}\ and\ \citenamefont
  {Manneville}(1980)}]{pomeau_80}%
  \BibitemOpen
  \bibfield  {author} {\bibinfo {author} {\bibfnamefont {Y.}~\bibnamefont
  {Pomeau}}\ and\ \bibinfo {author} {\bibfnamefont {P.}~\bibnamefont
  {Manneville}},\ }\bibfield  {title} {\enquote {\bibinfo {title} {Intermittent
  transition to turbulence in dissipative dynamical systems},}\ }\href@noop {}
  {\bibfield  {journal} {\bibinfo  {journal} {J. de Physique}\ }\textbf
  {\bibinfo {volume} {41}},\ \bibinfo {pages} {1235--1243} (\bibinfo {year}
  {1980})}\BibitemShut {NoStop}%
\bibitem [{\citenamefont {Ott}\ and\ \citenamefont {Sommerer}(1994)}]{ott_94}%
  \BibitemOpen
  \bibfield  {author} {\bibinfo {author} {\bibfnamefont {E.}~\bibnamefont
  {Ott}}\ and\ \bibinfo {author} {\bibfnamefont {J.~C.}\ \bibnamefont
  {Sommerer}},\ }\bibfield  {title} {\enquote {\bibinfo {title} {Blowout
  bifurcations: The occurrence of riddled basins and on-off intermittency},}\
  }\href@noop {} {\bibfield  {journal} {\bibinfo  {journal} {Phys. Lett. A}\
  }\textbf {\bibinfo {volume} {188}},\ \bibinfo {pages} {39--47} (\bibinfo
  {year} {1994})}\BibitemShut {NoStop}%
\bibitem [{\citenamefont {Saha}\ and\ \citenamefont {Feudel}(2008)}]{saha_18}%
  \BibitemOpen
  \bibfield  {author} {\bibinfo {author} {\bibfnamefont {Arindam}\ \bibnamefont
  {Saha}}\ and\ \bibinfo {author} {\bibfnamefont {Ulrike}\ \bibnamefont
  {Feudel}},\ }\bibfield  {title} {\enquote {\bibinfo {title} {Riddled basins
  of attraction in systems exhibiting extreme events},}\ }\href@noop {}
  {\bibfield  {journal} {\bibinfo  {journal} {Chaos}\ }\textbf {\bibinfo
  {volume} {28}},\ \bibinfo {pages} {033610} (\bibinfo {year}
  {2008})}\BibitemShut {NoStop}%
\bibitem [{\citenamefont {Benzi}\ \emph {et~al.}(1982)\citenamefont {Benzi},
  \citenamefont {Parisi}, \citenamefont {Sutera},\ and\ \citenamefont
  {Vulpiani}}]{benzi_82}%
  \BibitemOpen
  \bibfield  {author} {\bibinfo {author} {\bibfnamefont {Roberto}\ \bibnamefont
  {Benzi}}, \bibinfo {author} {\bibfnamefont {Giorgio}\ \bibnamefont {Parisi}},
  \bibinfo {author} {\bibfnamefont {Alfonso}\ \bibnamefont {Sutera}}, \ and\
  \bibinfo {author} {\bibfnamefont {Angelo}\ \bibnamefont {Vulpiani}},\
  }\bibfield  {title} {\enquote {\bibinfo {title} {Stochastic resonance in
  climatic change},}\ }\href@noop {} {\bibfield  {journal} {\bibinfo  {journal}
  {Tellus}\ }\textbf {\bibinfo {volume} {34}},\ \bibinfo {pages} {10--15}
  (\bibinfo {year} {1982})}\BibitemShut {NoStop}%
\bibitem [{\citenamefont {Pearson}\ and\ \citenamefont
  {Fox-{K}emper}(2018)}]{pearson}%
  \BibitemOpen
  \bibfield  {author} {\bibinfo {author} {\bibfnamefont {B.}~\bibnamefont
  {Pearson}}\ and\ \bibinfo {author} {\bibfnamefont {B.}~\bibnamefont
  {Fox-{K}emper}},\ }\bibfield  {title} {\enquote {\bibinfo {title} {Lognormal
  turbulence dissipation in global ocean models},}\ }\href@noop {} {\bibfield
  {journal} {\bibinfo  {journal} {Phys. Rev. Lett.}\ }\textbf {\bibinfo
  {volume} {120}},\ \bibinfo {pages} {094501} (\bibinfo {year}
  {2018})}\BibitemShut {NoStop}%
\bibitem [{\citenamefont {Isern-Fontanet}\ and\ \citenamefont
  {Turiel}(2021)}]{turiel}%
  \BibitemOpen
  \bibfield  {author} {\bibinfo {author} {\bibfnamefont {J.}~\bibnamefont
  {Isern-Fontanet}}\ and\ \bibinfo {author} {\bibfnamefont {A.}~\bibnamefont
  {Turiel}},\ }\bibfield  {title} {\enquote {\bibinfo {title} {On the
  connection between intermittency and dissipation in ocean turbulence: a
  multifractal approach},}\ }\href@noop {} {\bibfield  {journal} {\bibinfo
  {journal} {Journal of Physical Oceanography}\ ,\ \bibinfo {pages}
  {https://doi.org/10.1175/JPO--D--20--0256.1}} (\bibinfo {year}
  {2021})}\BibitemShut {NoStop}%
\bibitem [{\citenamefont {Smyth}(2020)}]{smyth_20}%
  \BibitemOpen
  \bibfield  {author} {\bibinfo {author} {\bibfnamefont {W.D.}\ \bibnamefont
  {Smyth}},\ }\bibfield  {title} {\enquote {\bibinfo {title} {Marginal
  instability and the efficiency of ocean mixing},}\ }\href@noop {} {\bibfield
  {journal} {\bibinfo  {journal} {J. Phys. Oceano.}\ }\textbf {\bibinfo
  {volume} {50}},\ \bibinfo {pages} {2141--2150} (\bibinfo {year}
  {2020})}\BibitemShut {NoStop}%
\end{thebibliography}%
\end{document}